# pH Regulates Ion Dynamics in Carboxylated Mixed Conductors

Zeyuan Sun,[1] Mengting Sun,[1] Rajiv Giridharagopal,[2] Robert C. Hamburger,[3] Siyu Qin,[1] Haoxuan Li,[1] Mitchell C. Hausback,[1] Yulong Zheng,[4] Bohyeon Kim,[1] Heng Tan,[5] Thomas E. Gartner III,[1*] Elizabeth R. Young,[3*] Christopher J Takacs,[6*] David S. Ginger,[2*] Elsa Reichmanis[1*]

[1]Department of Chemical and Biomolecular Engineering, Lehigh University, Bethlehem, PA 18015, United States

[2]Department of Chemistry, University of Washington, Seattle, WA 98195, United States

[3]Department of Chemistry, Lehigh University, Bethlehem, PA 18015, United States

[4]School of Chemistry and Biochemistry, Georgia Institute of Technology, Atlanta, GA 30332, United States

[5]Department of Computer Science and Engineering, Lehigh University, Bethlehem, PA 18015, United States

[6]Stanford Synchrotron Radiation Lightsource SLAC National Accelerator Laboratory, Menlo Park, CA 94025, United States

## Abstract

Coupled ionic and electronic transport underpins processes as diverse as electrochemical energy conversion, biological signaling, and soft adaptive electronics. Yet, how chemical environments such as pH modulate this coupling at the molecular scale remains poorly understood. Here, we show that the protonation state of carboxylated polythiophenes provides precise chemical control over ion dynamics, doping efficiency, solvent uptake and mechanical response. Using a suite of multimodal *operando* techniques and spectroelectrochemistry, supported by simulations, we reveal that pH dictates the balance between cation expulsion and anion uptake during electrochemical doping. Mapping across pH uncovers a quasi–non-swelling regime (≈pH 3–3.5) where charge compensation proceeds with minimal volumetric change yet pronounced stiffening. These findings establish molecular acidity as a general strategy to program ionic preference and mechanical stability, offering design principles for pH-responsive mixed conductors and soft electronic materials that couple ionic, electronic, and mechanical functionality.

## 1. Main

Electronic functionality has expanded far beyond computation and information storage to encompass intelligent, adaptive systems that engage directly with their environments including living organisms.[1] Bioelectronics, which bridges electronic and biological domains through devices capable of transducing electrical and biochemical signals bidirectionally, has emerged as a cornerstone of this evolution.[2,3] Such systems hold promise across diverse application spaces – from energy storage[4,5] and neuromorphic computing[6,7] to neurotransmitter regulation[8] and tissue-integrated sensing and actuation,[9,10] where ionic species serve as key mediators of signal and charge transport.

Among these ionic species, protons (H$^+$) occupy a uniquely central role. Local proton concentration (pH) governs charge transport in biological media and modulates fundamental processes, including neuronal signaling,[11,12] enzymatic catalysis,[13] and cellular metabolism.[14] Thus, pH is not a passive descriptor but an active regulatory parameter, dynamically shaping protein conformation,[15] enzyme activity,[16] and ion channel behavior.[17] Local gradients, whether at inflamed tissues,[18] tumor microenvironments[19] or neural synapses,[20] are tightly correlated with physiological and pathological states. Consequently, pH responsiveness is a fundamental building block for materials and devices to seamlessly interface with biological systems. Integrating such responsiveness into synthetic systems remains challenging. Unlike metal ions, protons often migrate through dynamic hydrogen-bonded networks, complicating transport modeling and experimental quantification.[21] Furthermore, proton-coupled electron transfer (PCET) processes are highly sensitive to local pH and hydration,[22] often leading to nonlinear, poorly understood redox behavior.[23] Disentangling the contributions of protons from coexisting cations remains a persistent hurdle. These challenges are compounded by limited availability of suitable materials and/or *operando* techniques capable of directly resolving pH-induced activity at buried interfaces. Thus, designing materials that enable selective, stable, reversible pH responsiveness in complex media requires deeper mechanistic understanding of coupled ionic/protonic-electronic dynamics across multiple length and time scales.

Organic mixed ionic–electronic conductors (OMIECs) provide a versatile platform for investigating coupled ionic–electronic transport across both biological and energy interfaces.[24] OMIECs can accommodate a broad range of ionic species—from physiological ions such as $Na^+$, $K^+$, and $Cl^-$ to energy-relevant species including $Li^+$, $TFSI^-$, and $PF_6^-$. Given the centrality of pH in regulating functionality, materials designed for bioelectronic or iontronic interfaces must tolerate and respond predictably to local pH, requiring molecular architectures capable of translating subtle acid-base equilibria into controlled ionic/electronic behavior. Such responsiveness has been reported in certain OMIECs, however, most studies interpret device-level performance shifts as indirect indicators rather than direct probes of underlying molecular mechanisms.[25–28] Elucidating how pH-induced structural and electrochemical changes evolve within complex ionic environments is essential to inform rational materials design and unlock new opportunities for engineering OMIECs for emerging applications.

Here, we demonstrate pH fundamentally reshapes the electrochemical doping mechanism in in p-channel carboxylated OMIECs, governing their doping efficiency, ion/water dynamics and structural evolution. The protonation/deprotonation equilibrium of the carboxylic acid side chains imparts finely tunable fixed charges with pH changes, while hole conduction along the conjugated backbone is balanced primarily by anion compensation. This configuration allows pH to act as a control that modulates ionic-electronic coupling without direct proton involvement, enabling pH-related process studies. Utilizing multimodal *operando* characterization techniques and course-grained (CG) molecular dynamics (MD) simulations, we reveal how pH regulates ion dynamics, leading to a cascade of structural, spectroscopic, and electrochemical transformations. Collectively, these findings establish carboxyl moieties as powerful design motifs, advancing the development of robust, pH responsive materials for next-generation bioelectronic and multi stimuli interfaces, and beyond.

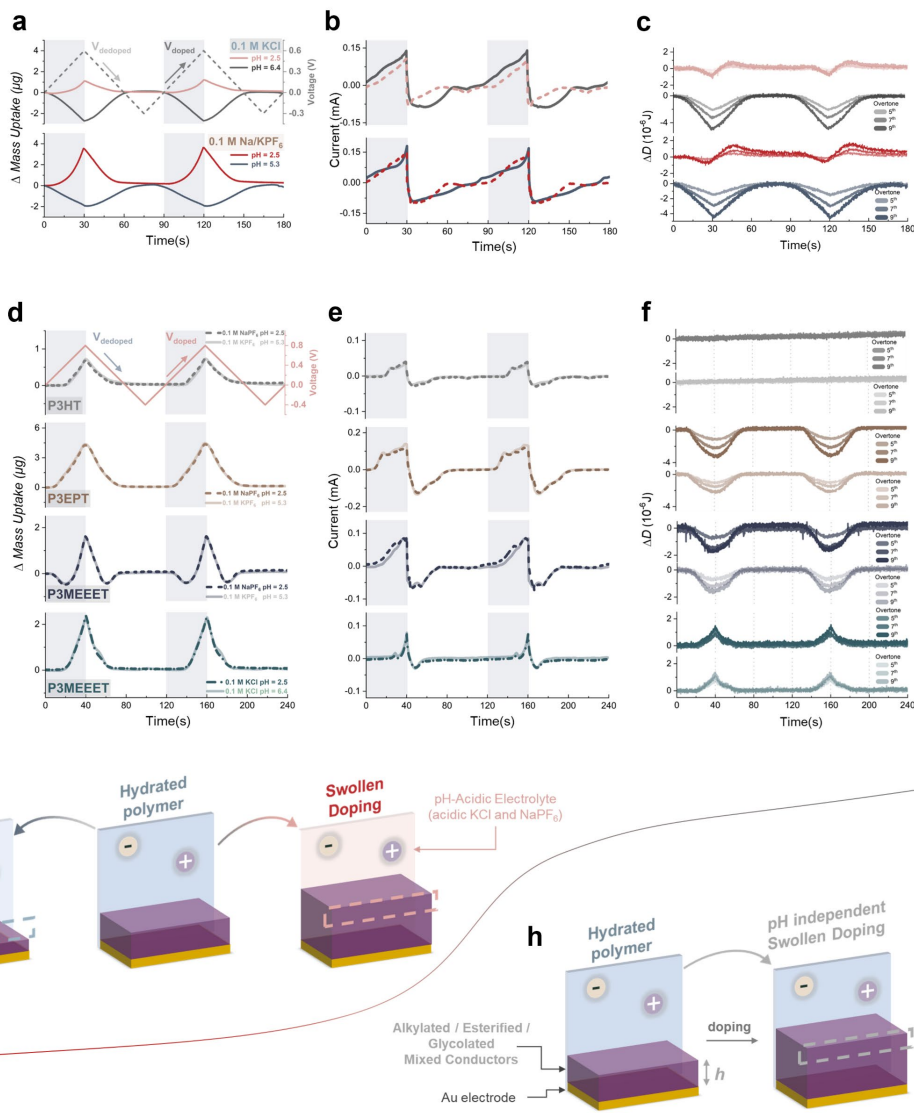

**Fig. 1**: **pH-dependent, and -independent mixed conduction in polythiophene-based OMIECs**. Chemical structures and pH-modulated electrochemical quartz crystal microbalance with dissipation (EQCM-D) measurements of the studied polymers. Measurements were conducted in both neutral and acidic aqueous electrolytes, including 0.1 M KCl (pH 6.4), 0.1 M KPF$_6$ (pH 5.3), 0.1 M KCl acidified to the designated pH with HCl, and 0.1 M NaPF$_6$, (pH 2.5). The **top panel** presents the **pH-dependent** group, featuring protonated carboxylic acid functionalized polythiophene (P3CBT-P) derived from as-cast P3KPT. Corresponding electrochemical measurements include: **(a)** mass change/voltage profiles (9$^{th}$ overtone), (b) cyclic voltammetry-derived current response, and **(c)** dissipation energy variation in neutral and acidic electrolytes. The **bottom panel** illustrates the **pH-independent** functional group, consisting of P3HT, P3EPT, and P3MEEET, with their corresponding **(d)** mass change/voltage profiles, (e) cyclic voltammetry-derived current response, and **(f)** dissipation energy variation in the same electrolytes as the carboxylated mixed conductors. **(g)** Schematic representation of the tuning principle for **pH-dependent mixed conduction**, where swelling behavior is modulated by pH—neutral pH favors deswollen doping, while acidic pH promotes swollen doping. **(h)** Schematic depiction of **pH-independent mixed conduction**, where swollen doping occurs irrespective of pH variations.

## pH dependent and independent mixed conduction

To probe the pH-responsive electrochemical behaviors, poly[3-(4-carboxylbutyl)thiophene-2,5-diyl] (P3CBT-P) served as a model carboxylated conjugated polyelectrolyte.[29,30] Electrochemical quartz crystal microbalance with dissipation monitoring (EQCM-D) measurements were conducted in both neutral and acidic aqueous electrolytes. As shown in **Fig. S1-S2**, neither the specific identity of the non-compensating ion (cations for p-channel doping) nor moderate variations in compensating ion concentration (anions in p-channel doping) alone account for the distinct electrochemical response, ruling out extrinsic factors and underscoring the intrinsic pH sensitivity of COOH functionality. Under neutral pH, EQCM-D reveals a net mass *loss* during electrochemical doping (**Fig. 1a**), consistent with prior observations of cation-carboxylate binding that drives deswelling as cation release outweighs anion uptake.[30] In contrast, acidic conditions elicit a dramatically different response, with pronounced mass *gain* during doping, reflecting a shift in ion exchange dynamics. We hypothesize that suppressed carboxyl dissociation at low pH weakens cation binding affinity, enabling anion uptake to dominate the swelling/gravimetric process.

Corresponding EQCM-derived voltammograms (**Fig. 1b**) reveal a pronounced shift in behavior, where acidic conditions delay onset of oxidation relative to neutral pH by ~0.2 V. Supporting evidence from passive swelling (**Fig. S3**) and contact angle measurements (**Fig. S4**) confirm pH-switchable wettability, where increased protonation reduces hydrophilicity, thus affecting ion-polymer interactions in an ion-independent manner. Notably, a higher degree of protonation leads to less swelling (~8% vs ~16%, respectively) under unbiased conditions.[31,32] Voltage-dependent dissipation profiles (**Fig. 1c**) provide further evidence of altered ion-polymer interactions, where acidic conditions produce a biphasic mechanical response – initial stiffening in the doping range followed by softening during dedoping – distinct from the reversible rigidity observed at neutral condition, indicating dynamic reorganization of the polymer network during doping. These results collectively point to pH-regulated modulation of polymer hydrophilicity, governed by the COOH protonation state and associated transient H-bonding and viscoelasticity.

CG-MD simulations of single P3CBT chains in aqueous KCl directly probed ion-polymer interactions as a function of sidechain protonation state (**Extended Data Fig.1a**). Simulations were performed in fully protonated (low pH) and fully deprotonated (high pH) states, (see Methods and Discussion 1). In direct correspondence with experimental hydrophilicity results above, deprotonated P3CBT showed extended configurations, while those of the protonated analog were collapsed, reflecting increased hydrophilicity in the deprotonated system. Radial distribution functions, $g(r)$, also support the above hypotheses around pH effects on polymer-ion and polymer-water interactions (**Extended Data Fig.1b-c**). Namely, deprotonated $COO^-$ groups promote sidechain-cation correlations and suppress backbone-anion correlations relative to COOH.

To establish that pH-responsive behavior is intrinsically linked to the presence of COOH moieties, we performed pH-modulated EQCM-D measurements on a series of polythiophene-based OMIECs, namely, poly(3-hexylthiophene-2,5-diyl) (P3HT), poly[3-(ethyl-5-pentanoate)thiophene-2,5-diyl] (P3EPT), and poly(3-[2-[2-(2-methoxyethoxy)ethoxy]ethyl]thiophene)−2,5-diyl) (P3MEEET) under $Cl^-$ and $PF_6^-$ electrolyte environments at varied pH (**Fig. 1d–e**). These *non*-ionizable polymers exhibited *no* appreciable variation in mass change, electrochemical response, or viscoelastic dissipation across pH environments, confirming that *ionizable*, COOH bearing side chains are essential for enabling proton-mediated doping dynamics. The striking transition observed in P3CBT-P, from deswollen doping state at neutral pH to highly swollen state under acidic conditions, highlights the criticality of pH in reconfiguring ion-polymer and polymer-polymer interactions, fundamentally altering the electrochemical doping mechanism (**Fig. 1g**). In contrast, non-ionizable analogues retain anion-dominated, pH-invariant swollen doping behavior (**Fig. 1h**). This pH responsive behavior appears robust across processing conditions: P3CBT-P and P3CBT (obtained from DMSO solutions of P3CBT) both exhibit pH-dependent mass uptake, electrochemical response and OECT performance (**Extended Data Fig.2**).

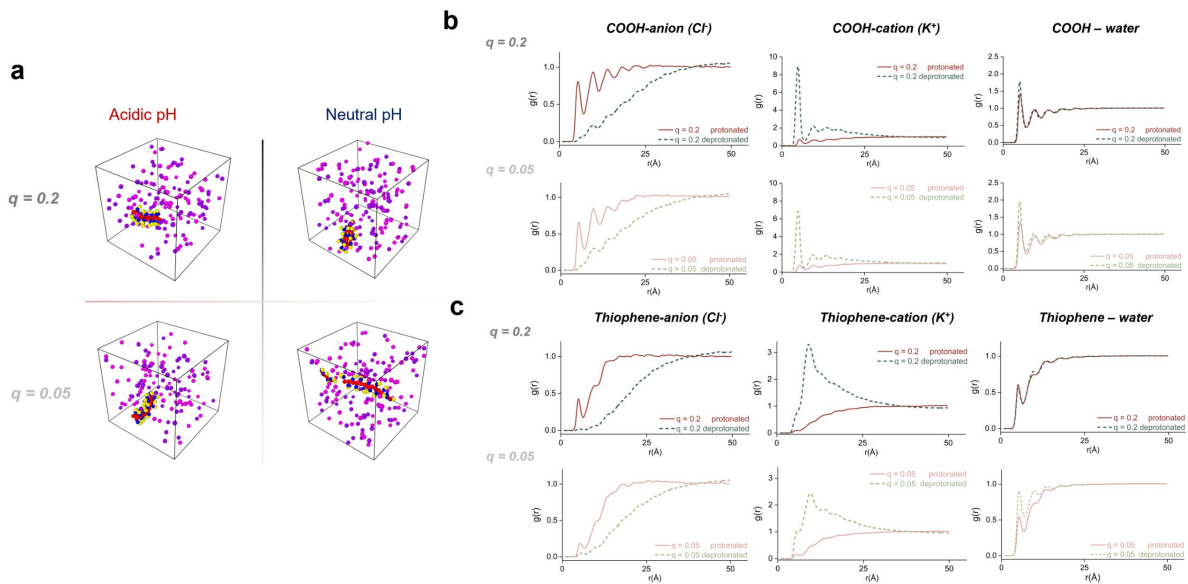

**Extended Data Fig. 1: a)** Coarse-grained molecular dynamics (MD) snapshots of a single P3CBT chain with different backbone charge densities (q = 0.2 and q = 0.05 per monomer) simulated in aqueous KCl under acidic (protonated) and neutral (deprotonated) pH conditions. Water molecules are hidden for clarity. **b)** Radial distribution functions (RDFs) between carboxylic acid groups and surrounding species (Cl$^-$, K$^+$, and water), and **c)** RDFs between thiophene backbones and the same species.

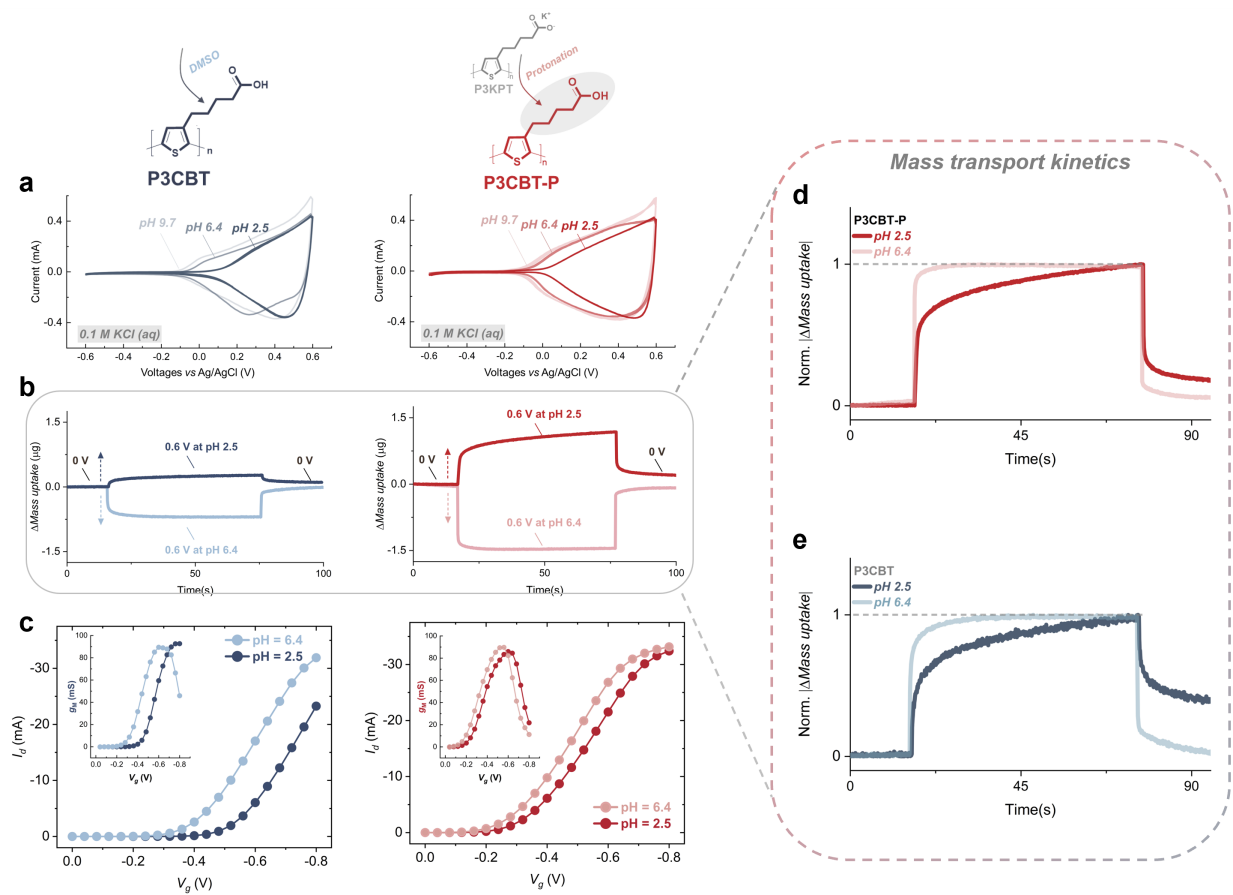

**Extended Data Fig. 2:** Side-by-side comparison of P3CBT (DMSO) and P3CBT-P (derived by protonation method) under varying pH conditions. **a)** cyclic voltammetry, **b)** real-time gravimetric behavior by EQCM-D, and **c)** interdigitated OECT (iOECT) transfer curves in 0.1 M KCl at neutral and acidic pH. Alkaline condition in a) were not further explored due to complications arising from acid-base neutralization and instability issue that potentially dissolve the polymers. Normalized mass transport kinetics (from b) during electrochemical equilibrium, highlighting differences in mass transport kinetics at different pH conditions for **d)** P3CBT-P and **e)** P3CBT, respectively. Notably, the degree of COOH protonation strongly influences electrochemical behavior, with earlier onset potentials observed under highly deprotonated (pH 6-9) *vs.* highly protonated (pH 2.5) state. Mass transport kinetics reveal a consistent trend across both polymers with gravimetric response that is slower under pH acidic conditions. Despite reduced swelling associated with DMSO processing, the results align with increased hydrophobicity and altered doping dynamics in the protonated state. Together, these comparisons highlight the unique pH-sensitivity imparted by COOH and establish a design strategy for modulating mixed conduction *via* control of local proton concentration.

## Real-time Probing of Ionic and Structural Reorganization with *Operando* X-ray Scattering and Scanning Probe Techniques

To further explore the impact of pH on ion dynamics, *operando* grazing incidence X-ray fluorescence (GIXRF) enabled tracking of ion movement in and out of the bulk thin film during electrochemical cycling. Analysis focused on the potassium K-edge emission (~3.4 keV), a direct probe of cation concentration in the carboxylated polythiophenes.[30] All K$^+$ signals were normalized to the elastic scattering peak, enabling direct qualitative comparison between environments (**Fig. S5-S6**). GIXRF spectra (**Fig. 2a**) show pH-dependent potassium intensity modulation. At pH 6.4, the signal exhibits pronounced cyclic oscillations, indicating reversible expulsion/reinsertion of K$^+$ during (de)doping—consistent with a cation-compensated mechanism facilitated by deprotonated COOH groups. In contrast, at pH 2.5, the K$^+$ signal is weaker with reduced magnitude, suggesting diminished cation involvement and a shift toward anion-dominated doping. Notably, after dedoping at pH 6.4, the K$^+$ intensity increases, indicating net cation accumulation within the polymer, a trend not observed under acidic conditions. These observations demonstrate that the COOH protonation state governs cation transport, even within the highly swollen, acidic regime. Complementary liquid-phase atomic force microscopy (AFM) with bias applied *in situ* under identical pH conditions (**Fig.2b**) corroborates coupling between mass and dimensional change: doping at neutral pH yields a gradual thickness *decrease* up to ~28 nm (**−8%**), whereas acidic conditions produce a gradual *increase* (~37 nm, **+15%**) (**Fig. S7**). These results confirm the hypothesis that reduced cation-polymer binding at low pH drives a transition toward anion-dominated, swollen states.

Further evidence of stiffness correlated with electrochemical strain was demonstrated *in situ* (**Fig. 2c**), where COOH functionalization shows a lower elastic modulus (~520 MPa) in acidic *vs.* neutral (~800 MPa) pH at 0 V, likely owing to complex protonation-hydration interactions.[33,34] Upon doping (-0.8 V), both pH conditions lead to pronounced stiffening, with the elastic moduli increasing to 800 MPa at pH 2.5, and ~1250 MPa at pH 6.4. This trend is consistent with the reduced dissipation observed in **Fig.1c** and is reflective of doping-induced structural and electrostatic effects. In both cases, doping injects anions while expelling cations previously associated with fixed COO$^−$ groups. Ion exchange reduces internal hydration, collapses free volume and eliminates the plasticizing effect.[35] Plausibly, pH-modulated ionic crosslinking during doping outweighs plasticization,[36] and additional anion-COO$^-$ repulsion may further restrict chain mobility, enhancing stiffness. These effects transform doping from a softening mechanism[37] (common in glycolated polymers), into a stiffening one, highlighting a distinct structure-mechanics coupling in carboxylated OMIECs.

To further study the bulk thin film ion content structural dynamics in the pH induced changes, we use *operando* grazing incidence wide-angle X-ray scattering (GIWAXS) with simultaneous GIXRF collection (**Fig. 2d**). Although P3CBT-P remains the benchmark carboxylated mixed conductor across the series, the longest side chain carboxylhexyl variant, P3CHT, exhibits well-defined *ex-situ* GIWAXS feature (**Figure S8-S9**). This material retains pH-dependent swelling response (**Extended Data Fig.3**), making P3CHT ideal for *operando* GIWAXS-GIXRF studies. **Fig. 2e** summarizes the evolution of P3CHT elemental and structural response during CV at pH 6.4 and 2.5 in 0.1 M KCl. GIXRF reveals reversible K$^+$ insertion/expulsion. The GIWAXS lamellar (100) reflection remains nearly unchanged at pH 6.4 ($\Delta d \approx 0.6$ Å, 0.3%) but is pre-expanded by ~3.1 Å (15%) at pH 2.5, indicating substantial ionic intercalation prior to cycling. Subsequent doping at pH 6.4 induces minimal expansion (~0.5 Å), consistent with K$^+$ expulsion (deswollen doping), whereas pH 2.5 exhibits an ~1.2 Å expansion and pronounced structural hysteresis aligned with electrochemistry delay. The π–π (010) peak contracts slightly under both conditions (< 0.1 Å), suggesting tighter yet intact backbone packing. These results indicate highly anisotropic electrochemical swelling, confined primarily to side-chain lamellae, while the conjugated cores remain structurally and electronically coupled throughout.

Together, *operando* GIWAXS-GIXRF results reveal a coherent picture of pH-dependent ion distribution in carboxylated OMIECs. In crystalline domains, the behaviors of cation/anion pairs exhibit pronounced pH dependence, with anion playing the dominant structural role. While doping induces lamellar expansion and π–π contraction, the extent of expansion is minimal at neutral pH but pronounced under acidic conditions. Protons are likely synchronized with cation motion under acidic conditions.[38] Correlating this trend with macroscopic swelling reveals that ionic species primarily occupy amorphous domains, where pH-dependent, spatially heterogeneous volumetric change occurs.[39] This localization rationalizes the coupled structural and mechanical responses, indicating that the amorphous matrix governs volumetric dynamics, while crystalline order sustains electronic transport.

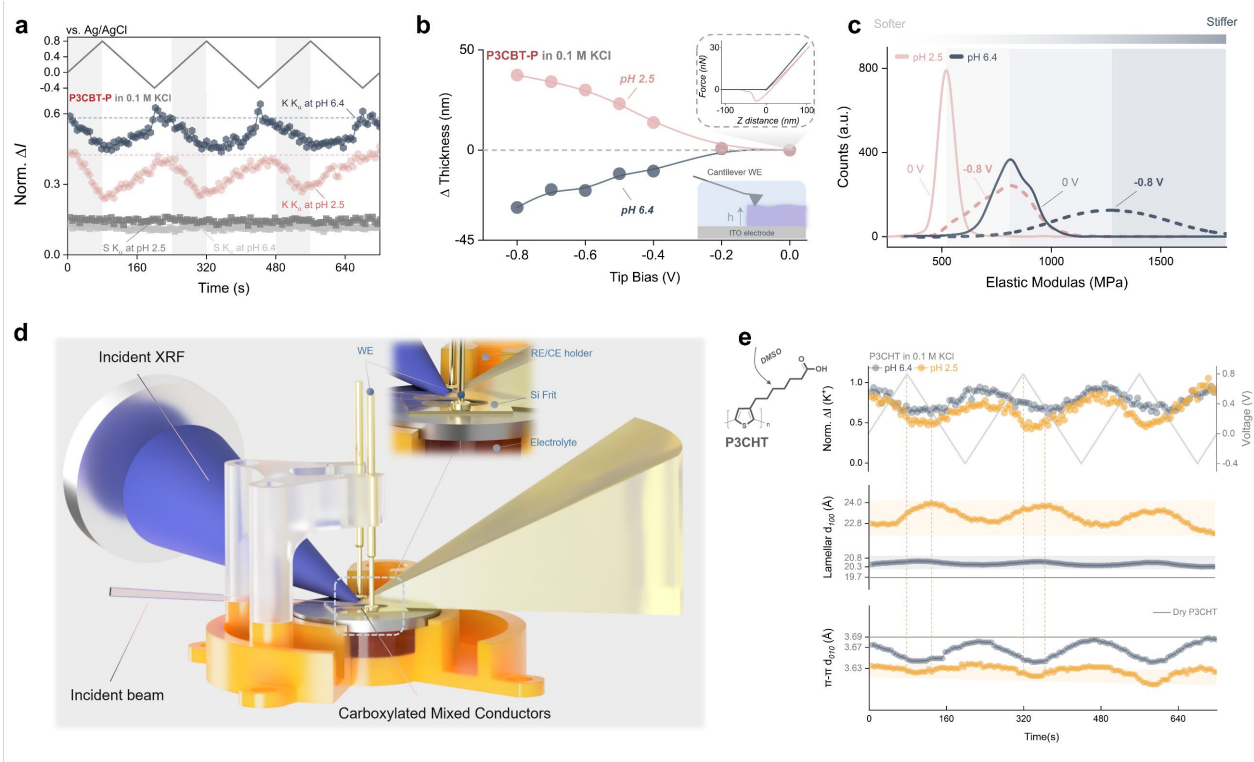

**Fig. 2. Multimodal *operando* characterization of carboxylated polythiophenes**. a) Operando grazing incident X-ray fluorescence (GIXRF) of P3CBT-P in 0.1 M KCl at pH 6.4 and pH 2. b) In situ AFM thickness tracking and AFM nanoindentation during 0 V potential holds at pH 6.4 and pH 2.5 c) in situ force-mapping under the same electrolyte and pH conditions. d) Combined operando grazing- incidence wide-angle X-ray scattering and X-ray fluorescence (GIWAXS+GIXRF) of P3CHT in 0.1 M KCl at pH 6.4 and pH 2.5.

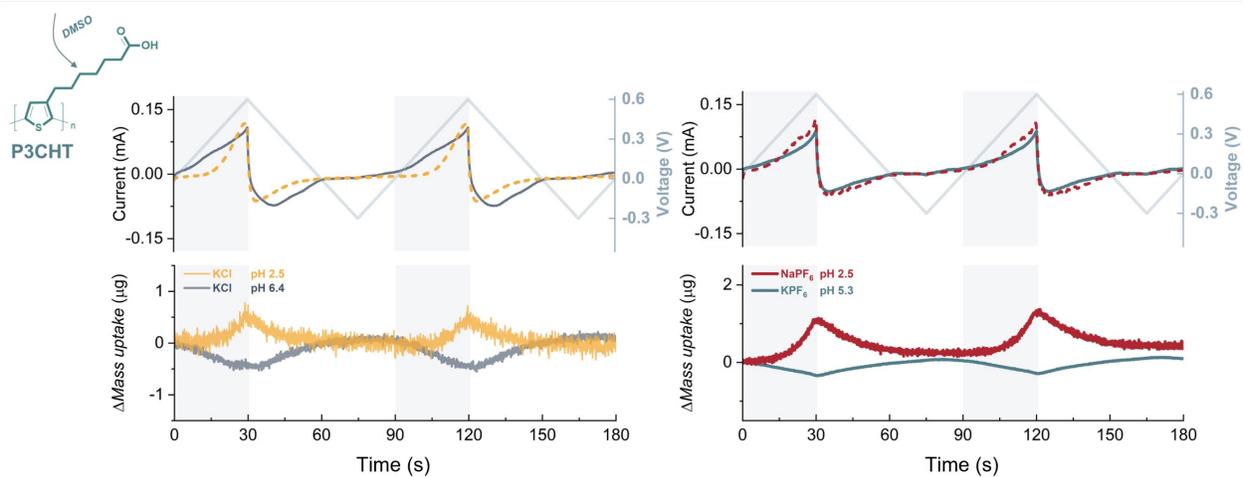

**Extended Data Fig. 3**: Side-by-side comparison of P3CHT (DMSO) under different pH conditions in chloride-based (left) and PF$_6^-$-based (right) electrolytes. Both cases exhibit clear pH-dependent swelling behavior, indicating that P3CHT can serve as a representative model for carboxyl-functionalized polythiophenes in GIWAXS and GIXRF studies

# Protonation-Mediated Modulation of Charge Delocalization and Doping

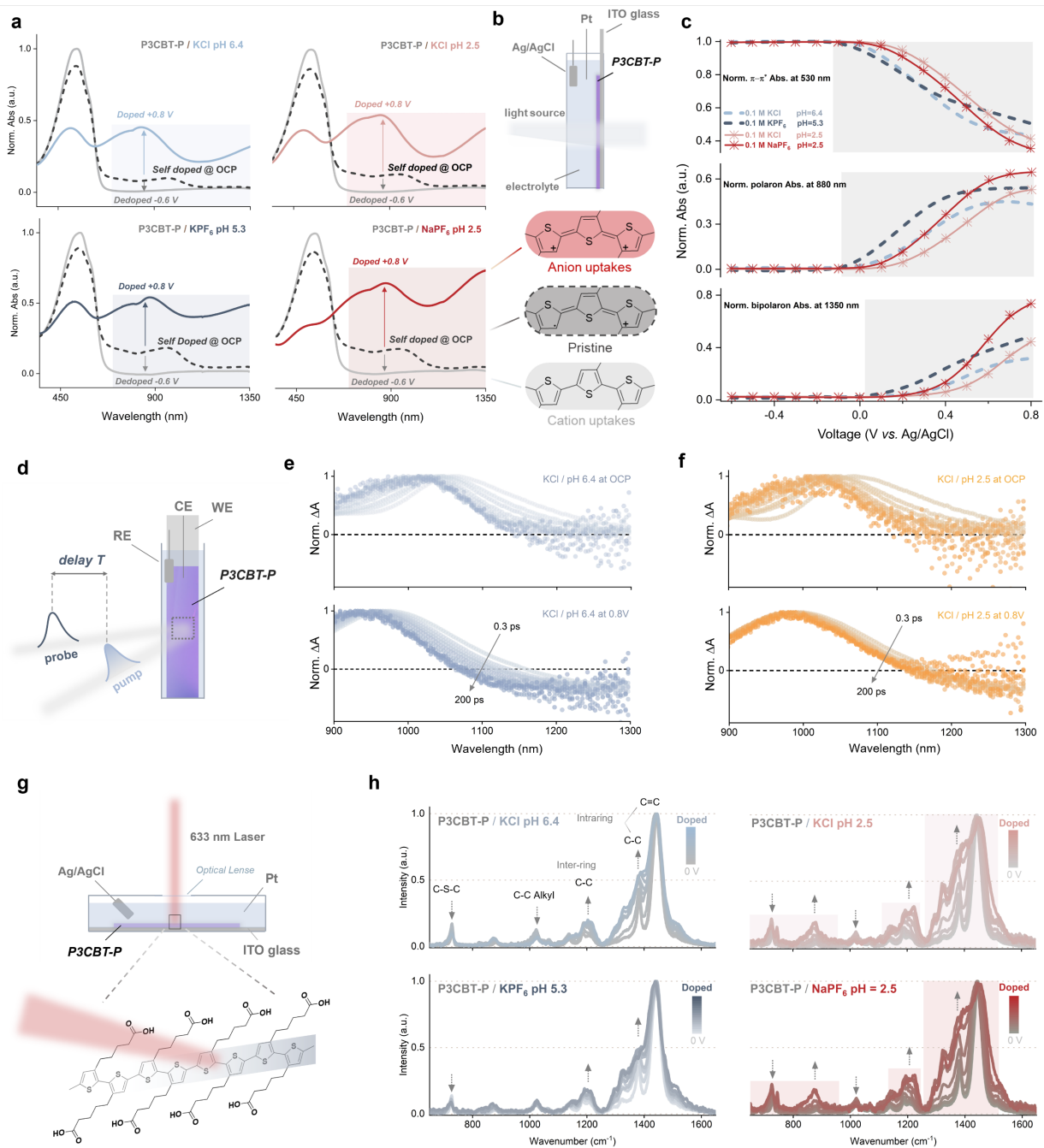

**Fig. 3**: **pH-dependent photophysical characterization of P3CBT-P**. a) Spectroelectrochemistry of P3CBT-P in pH neutral (0.1 M KCl, pH 6.4; 0.1 M KPF$_6$, pH 5.3) and pH acidic (0.1 M KCl, pH 2.5; 0.1 M NaPF$_6$, pH 2.5) electrolytes. b) Schematic of the *in situ* spectroelectrochemical cell. c) the evolution of steady-state absorbance during electrochemical doping from -0.6 V to 0.8 V (vs. Ag/AgCl), showing changes in the at 0-1 neutral (530 nm), polaron (880 nm), and bipolaron (1350 nm) peaks. Figure 3.d) Schematic of the *operando* transient absorption spectroscopy (TAS) setup, and the doped-state (0.8 V) TAS spectra at e) pH neutral, and f) pH acidic KCl condition at open circuit potential and doping potential

0.8 V vs Ag/AgCl. g) *operando* Raman experiment setup, and h) evolved spectrum for P3CBT-P in neutral (0.1 M KCl, pH 6.4; 0.1 M KPF$_6$, pH 5.3) and acidic (0.1 M KCl, pH 2.5; 0.1 M NaPF$_6$, pH 2.5) electrolytes from 0 V to 0. 8 V doped condition

      To elucidate the pH-dependent modulation of charge carrier dynamics further, *operando* ultraviolet-visible-near-infrared (UV-Vis-NIR) spectroelectrochemistry was performed on P3CBT-P (**Fig. 3 a-b**). Across all electrolyte conditions, discrete electronic states are revealed: a neutral-like configuration under reducing bias (-0.6 V), a doped state under oxidative bias (+0.8 V), and an intermediate "self-doped" state at open circuit potential (OCP).[40,41] This OCP feature reflects a pre-existing population of charge carriers and/or ionic species within the polymer, suggesting intrinsic predoping behavior in the absence of applied bias (**Fig. S10**).[42] Upon anodic polarization, systematic attenuation of the ~530 nm absorption is accompanied by emergence of polaron (~880 nm) and bipolaron (~1350 nm) bands, consistent with oxidative doping and the progressive delocalized charge carrier formation. Voltage-dependent spectral evolution (**Fig. 3c**) exhibits a clean, monotonic trend, underscoring electrochemical control over charge carrier formation. Acidic electrolytes induce a noticeable delay in polaron and bipolaron growth onset vs. neutral conditions: Cl$^-$-based systems show a more pronounced lag than PF$_6^-$. Nevertheless, at higher potentials (≥ 0.7 V), films doped under acidic pH exhibit elevated absorption intensities across all doping bands, indicating a higher final doping level. These results suggest that pH governs ion exchange kinetics and carrier formation, enabling access to deeper doping regimes in carboxylated polythiophenes—potentially *via* enhanced anion uptake with reduced electrostatic repulsion.

      To further probe the nature of photoexcited charge carriers under varying pH conditions, *operando* femtosecond transient absorption spectroscopy (TAS) was performed. In neutral KCl, the spectra exhibit a well-defined photoinduced absorption (~1000 nm), with an onset near 940 nm (**Fig. S11-S12**). The dynamics show rapid decay components on the sub-nanosecond timescale (**Fig. S13**, ~0.1-0.2 ns), consistent with polaronic relaxation processes in moderately doped conjugated system (**Fig. 3e**).[43] At acidic pH, TAS reveals a broader, red-shifted band, accompanied by similarly fast decay. Generally, the red-shifted and faster-decaying polaron spectra in doped systems indicate more delocalized charge carriers, in agreement with GIWAXS evidence of reduced π–π stacking distance at lower pH (**Fig. 3f**).

      *Operando* Raman spectroscopy explored molecular-level structural changes underlying pH-modulated doping (**Fig. 3g**). Across all conditions, a progressive evolution of vibrational features is observed upon doping (**Fig. 3h**), notably in the C=C stretching regions, reflecting changes in backbone conjugation and planarity.[44] As the applied potential increases, the quinoid-like C=C symmetric stretch intensity grows significantly, accompanied by a redshift with significantly higher $I_{C-C}/I_{C=C}$ under lower pH indicating increased planarization and electronic delocalization.[29,45] At lower wavenumbers (700-1000 cm$^{-1}$) associated with ring deformation and side chain interactions, pH-dependent differences emerge. Under acidic conditions, new bands appear, suggesting formation of distinct structural motifs or ion-polymer interactions – possibly arising from enhanced anion uptake or stronger electrostatic coupling between protons and carboxylate groups. Collectively, the UV–Vis–NIR, TAS, and Raman results reveal that pH modulation both tunes the doping level and reshapes charge carrier electronic configuration and relaxation dynamics. These *operando* spectroelectrochemical insights establish that pH-regulated ion exchange governs carrier delocalization and overall optoelectronic character of carboxylated polythiophenes. Taken together, a complete mechanistic picture emerges, revealing pH-dependent ion preference and dynamics in carboxylated OMIECs as shown in **Extended Data Fig.4** (Discussion 2, **Table S4, Fig. S14**).

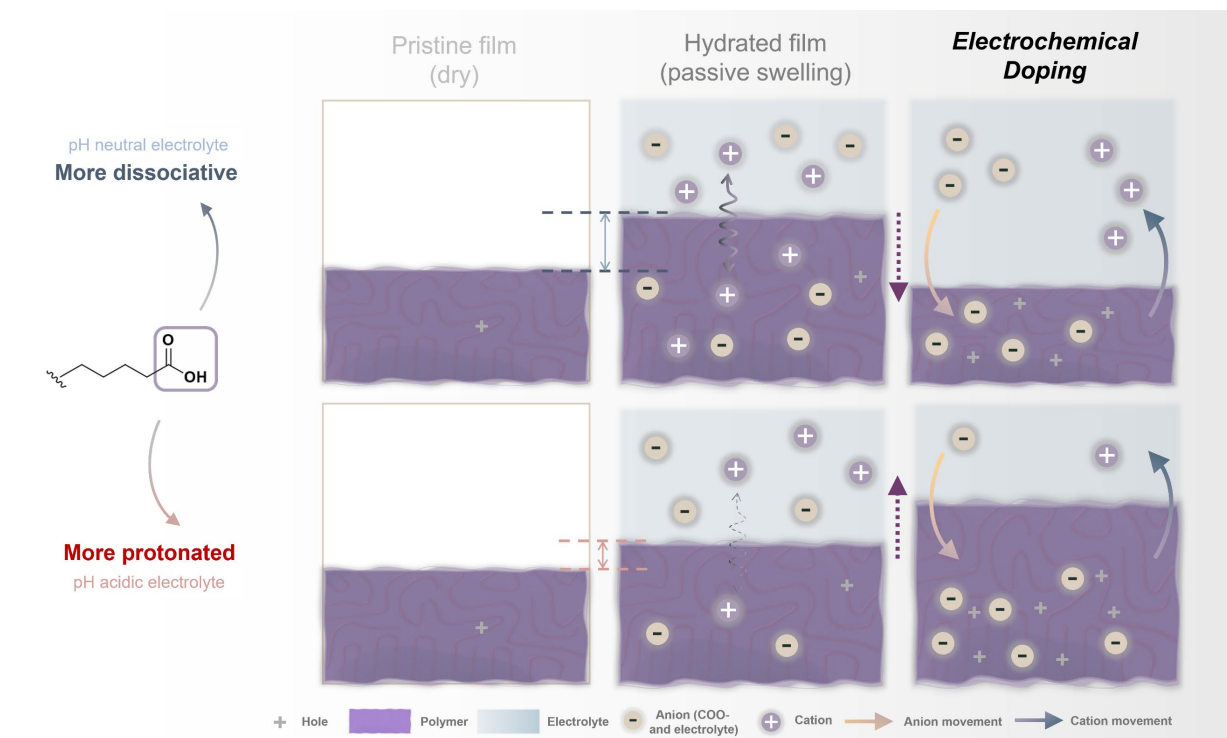

**Extended Data Fig. 4**: Schematic of the pH-dependent ion dynamics in carboxylated mixed conductors. Carboxylic acid functionalization introduces a small density of intrinsic charge carriers even without applied bias. At neutral pH, deprotonation of carboxylic acid groups generates negatively charged –COO⁻ sites that preferentially attract cations from the electrolyte while partially excluding anions. Under acidic conditions, protonation reduces the population of –COO⁻ groups, leading to weaker cation association but increased anion uptake. (see Discussion 2, Table S4, Fig. S14, and CG-MD results from Extended Fig. 1) During electrochemical doping, cation expulsion outweighs anion uptake, yielding a deswollen doping mode with a limited doping level. Under acidic conditions, protonation reduces the fixed charge density, weakening cation affinity while allowing greater anion ingress. The diminished Donnan exclusion at low pH leads to a swollen doping mode, reaching higher oxidation states at elevated bias.

## Macroscale and Microscale Insights into Quasi–Non–Swelling Behavior at the Critical pH in Carboxylated Mixed Conductors

As the pH-dependent behavior was consistently confirmed across *operando* measurements, we extended the investigation to finer pH resolution. P3CBT was characterized in 0.1 M KCl across a range from neutral to increasingly acidic conditions (**Fig. 4a-b**). Pronounced deswelling occurs upon doping at pH 5.5; while decreased pH leads to progressive reduction in mass change, indicating that increased protonation suppresses deswelling. A critical point emerges near pH 3.0-3.5 where gravimetric response becomes minimal: mass uptake decreases by approximately 85% and 95%, respectively, for thin and thick films. To verify this quasi-non-swelling state, we performed *in-situ* AFM with an applied bias under the same conditions, revealing only ~2.5 nm thickness change during doping – corresponding to less than 0.7% of the initial film thickness (**Fig. 4c**). A similar effect was observed in P3CBT-P at pH 3.5 (**Extended Data Fig.**

**4a-b**), where the thickness change was limited to <0.3% (**Fig.S15**), further supporting the presence of a volumetrically stable, nanoscopic doping mode at the critical pH.

These findings establish that a quasi-non-swelling doping mode can be achieved at a critical pH for carboxylated OMIECs. This volumetrically stable condition likely reflects a charge-compensation regime where the number of hydrated anions entering the film is balanced by hydrated cations and protons exiting, minimizing macroscale structural expansion despite ongoing redox activity (**Fig. 4d**). This state is linked to the protonation equilibrium of the COOH groups, suggesting that it arises from an electrochemically balanced dissociative regime unique to COOH-functionalized polymers. The precise pH at which this balance occurs, however, is influenced by environmental factors, including ion identities, electrochemical window and polymer effective pKa, which modulate the onset of volumetric stability.

Despite achieving volumetric stability at pH ~3.5, *operando* force measurements reveal substantial micromechanical changes during electrochemical doping (P3CBT and P3CPT-P, **Fig. S16**). A pronounced increase in elastic modulus occurs upon doping, indicating that the polymer backbone and side chains experience localized mechanical changes even absent measurable swelling. This contrast between macroscale stability and microscale mechanical response highlights a distinct mesoscale regime in which charge redistribution and ion compensation drive significant internal reorganization without large-scale dimensional expansion. These findings suggest that quasi-non-swelling behavior does not equate to mechanical passivity, but rather reflects a finely balanced state where internal electrostatic forces and local conformational rearrangements are accommodated within geometrically constrained polymer matrices. Overall, this behavior confers a key advantage in controlling swelling, positioning carboxylated OMIECs as promising candidates for mechanically robust devices.

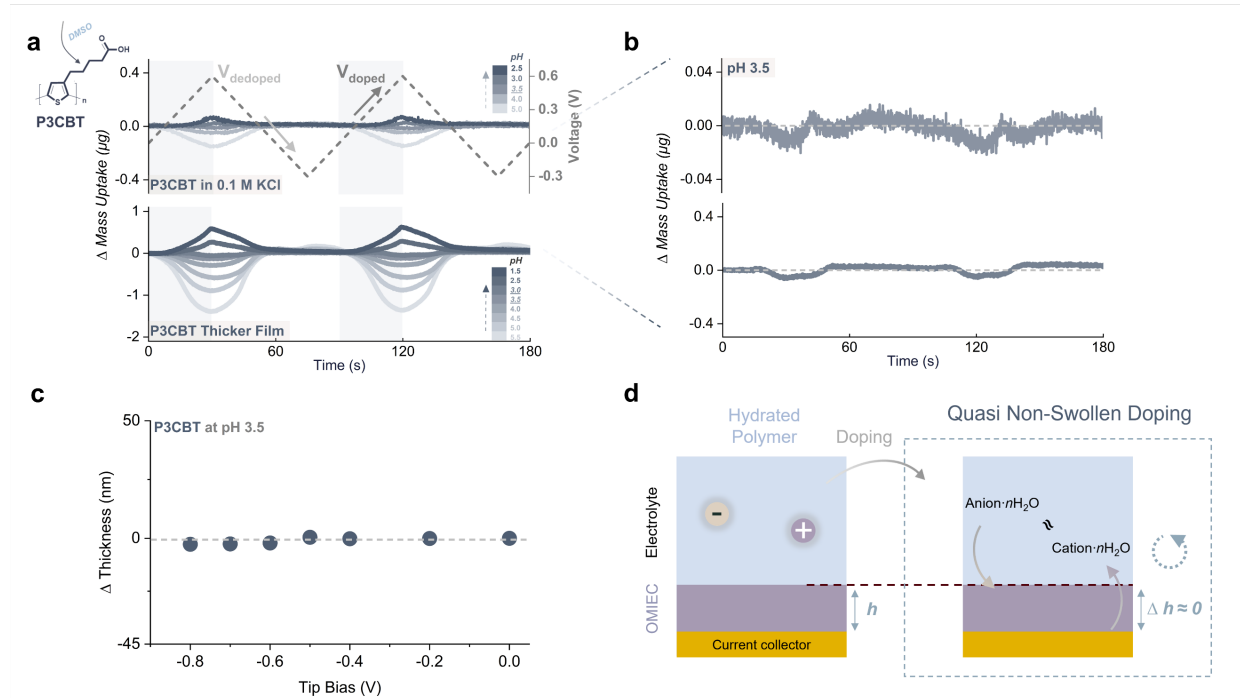

**Fig. 4**: **Critical pH enables quasi-non-swelling behavior for carboxylated mixed conductors**. a) pH-dependent gravimetric response of P3CBT in 0.1 M KCl, spanning pH neutral to acidic conditions. b) magnified view near the critical pH of 3.5, under two thickness, revealing a non-monotonic swelling trend;

c) in situ AFM confirming volumetrically-stable thickness under doping at the critical pH. d) schematic illustration of the quasi-non-swelling mechanism.

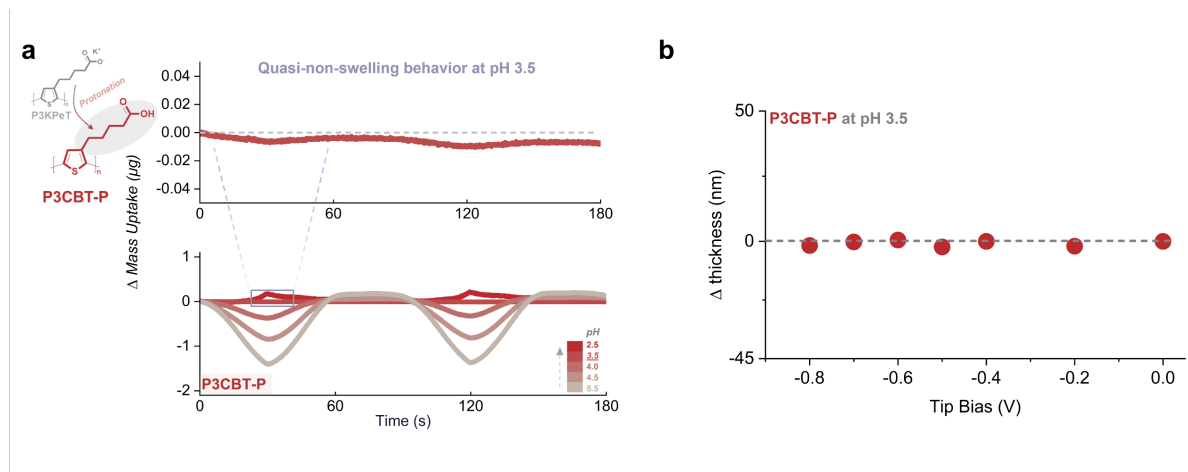

**Extended Data Fig. 5**: **a)** pH-dependent gravimetric response of P3CBT-P in 0.1 M KCl, highlighting the quasi-non-swelling behavior near the critical pH ≈ 3.5. The non-monotonic swelling trend mirrors that in Fig. 4, confirming that the quasi-non-swelling regime is process-independent across solvents and governed by the carboxylic acid functionalization. **b)** In-situ AFM further confirms volumetrically stable film thickness under electrochemical doping at the critical pH.

## 2. Outlook

Using a suite of *operando* multimodal characterization techniques, we reveal a pH-modulated doping mechanism governing charge compensation in carboxylated mixed conductors. The observed pH-dependent ion distributions originate from the intrinsic side chain chemistry, whose tunable dissociation–association equilibrium dictates overall ion dynamics. This coupling between chemical speciation and electrochemical behavior manifests across multiple length scales—from mesoscale (de)swelling to nanoscale structural rearrangements—coherently modulating the material's electrochemical, optoelectronic, and mechanical responses. By probing these dynamics at finer resolution, we identify a quasi-non-swelling regime near the critical pH, where electrochemical doping proceeds with nearly no volumetric change. This behavior underpins improved structural reversibility and suggests pathways toward enhanced long-term stability and biointerface compatibility, paving the way for realistic implantable and soft electronic applications.

While PCET typically governs n-channel OMIECs—where protons act as mobile compensating species—the mechanism in carboxylated p-channel polymers is distinct. Here, protons serve not as charge carriers but as chemical modulators of the fixed charge landscape, tuning side-chain deprotonation and thereby regulating mixed conduction indirectly. This decoupling of proton transport from electrochemical doping enables controlled, environment-responsive operation and unveils a unified framework for ionic–electronic–mechanical coupling in soft semiconductors. The ability to regulate ionic and volumetric responses through molecular acidity offers routes to electrochemical actuators with reversible deformation, while the sensitivity of structural and electronic states to pH and doping provides opportunities for light-responsive or photochemically coupled systems. Together, these insights establish a molecular-acidity-driven design paradigm for organic mixed conductors that seamlessly integrates electronic, mechanical, and photonic functionality for next-generation adaptive materials.

## 3. Method

Polymer Solution and Film Preparation: P3K(Pe)T (poly(3-potassium-5-pentanoate)thiophene-2,5-diyl), Mw = 28 kDa, PDI = 2.0, Batch # BLS26-51), P3K(He)T (poly(3-potassium-6-hexanoate) thiophene-2,5-diyl), Mw = 22 kDa, PDI = 1.8, Batch # PTL34-75), P3K(Hp)T (poly(3-potassium-7-heptanoate) thiophene-2,5-diyl), Mw = 65 kDa, PDI = 2.6, Batch # BLS25-60), P3CBT (poly(3-(4-carboxylbutyl)thiophene-2,5-diyl), Mw = 39 kDa, PDI = 2.3, Batch # PTL39-74), P3CPT (poly(3-(4-carboxylpentyl)thiophene-2,5-diyl), Mw = 22 kDa, PDI = 1.8, Batch # BLS26-41), P3CHT (poly(3-(4-carboxylhexyl)thiophene-2,5-diyl), Mw = 65 kDa, PDI = 2.6, Batch # PTL39-70), P3EPT (poly(3-(ethyl-5-pentanoate)thiophene-2,5-diyl), Mw = 39 kDa, PDI = 2.3, Batch # PTL37-60), P3HT (Poly(3-hexylthiophene-2,5-diyl), Mw = 74 kDa, PDI = 2.2, Batch # PTL39-03). P3MEEET (Poly(3-[2-[2-(2-methoxyethoxy)ethoxy]ethyl]thiophene)−2,5-diyl), Mw = 18 kDa, PDI = 1.6, Batch # BLS26-79),) were purchased from Rieke Metals Inc. *p*-Toluenesulfonic acid monohydrate (pTsOH-H2O, 98%, Sigma-Aldrich), potassium chloride (99.9%, Sigma-Aldrich), potassium hexafluorophosphate (99.9%, Sigma-Aldrich), sodium hexafluorophosphate (99.9%, Sigma-Aldrich), dimethyl sulfoxide (99.9%, Sigma-Aldrich), 1,2-dichlorobenzene (99%, Sigma-Aldrich), acetone (99.5%, Sigma-Aldrich), anhydrous chloroform (99.9%, Sigma-Aldrich), methanol (99.8%, Sigma-Aldrich), and isopropyl alcohol (99.5%, Sigma-Aldrich and used as received. Deionized water with a resistivity of 18.2 mΩ cm was obtained using an Aries water purification system. All protonated films, including P3CBT-P, P3CPT-P and P3CHT-P were fabricated following previously reported acidification methods.[29,30,46] by dissolving carboxylate salt into water at a concentration of 2 mg mL$^{-1}$ followed by deposition and acidification to COOH. P3CBT, P3CPT and P3CHT (2 mg) was dissolved in DMSO (1mL); P3EPT (2mg) was dissolved in 1,2-dichlorobenzene (1 mL), and P3MEEET (2mg) was dissolved in chloroform (1 mL). All polymer solutions were stirred at room temperature overnight. They were then spray-cast using a gravity feed Iwata Eclipse HP-CS airbrush, and heated to 60°C for P3MEEET, 80°C for P3CBT-P and 130°C for P3CBT and P3EPT. P3HT was dissolved in chloroform at 55°C under stirring for 30 min, then left to cool to room temperature and then spin coated at a speed of 1500 rpm for 60 s. The films were cast onto various substrates including glass, indium tin oxide (ITO) coated glass (Delta Technologies, resistivity = 8–12 Ω sq$^{-1}$) for *in situ* measurements, gold sensors for quartz crystal microbalance with dissipation (QCM-D) (Quartz PRO, QCM sensor, active area of 1.13 cm$^2$), and silicon for grazing incidence wide-angle X-ray scattering (GIWAXS) characterization.

Contact Angle and Thickness Measurement: Contact angle measurements were performed on a ramé-hart Model 260 Standard Contact Angle Goniometer/Tensiometer with DI water and electrolyte on polymer-coated glass. Thickness measurements were performed using an 3D optical profilometer/interferometer system ZeGage™ Pro.

Cyclic voltammetry (CV): CV (scan rate of 50 mV s$^{-1}$ with a step size of 2 mV) was conducted using an AMETEK PMC 200 Potentiostat/Galvanostat and a three-electrode setup. The WE was prepared by spray-coating the polymer films onto ITO-coated glass slides (Delta Technologies, resistivity = 8–12 Ω sq$^{-1}$). A platinum counter wire was used as the CE, and a standard Ag/AgCl electrode (3 M aqueous KCl inner solution, BASi) was used as the RE. 0.1 M KCl (*aq*) and K/NaPF$_6$ (*aq*) solution was used as the electrolyte. To ensure accurate measurements, all electrolytes were degassed under argon flow (15 min) prior to and during the measurement process. pH values were accurately measured using an InLab Micro Pro-ISM pH sensor (Mettler-Toledo) interfaced with a Seven2Go pH/ion meter (Mettler-Toledo).Quartz Crystal Microbalance with Dissipation Monitoring (QCM-D): Passive swelling measurements were conducted with a Q-Sense Explorer Analyzer First, the response of the bare Ti/Au sensors were recorded using QCM-D in air condition, followed by measurements after injection of 0.1 M KCl (*aq*) electrolyte into the chamber. These control measurements resulted in significant shifts in frequency and dissipation due to density differences between the media, which were excluded from the swelling percentage calculation. The sensors were then removed, and the conjugated polymer films were spray coated directly onto the same sensors for P3CBT and P3EPT, additional acidification steps are performed to make P3CBT-P thin film as previously described.[29,30,46] The absolute difference in frequency between the bare sensor and the Ti/Au/Polymer coated sensors was compared using the "stitched data" function of Q-Soft software. This function accounted for density differences and allowed for direct determination of mass changes per unit area using the Sauerbrey equation (**Equation 1**). The calculated mass changes were then converted to thickness changes, considering the sensor area and assuming a density of 1 g-cm$^{-3}$ for the polymers in different states (dry and wet).

$$\frac{\Delta m}{A} = \frac{-17.7}{n}\Delta f_n$$ **Equation 1**

Electrochemical Quartz Crystal Microbalance with Dissipation Monitoring (EQCM-D): Active swelling measurements under electrochemical doping/dedoping were performed using a Gamry interface 1010B coupled with Q-sense electrochemistry module (QEM401, Biolin Scientific). The three-electrode setup comprised an Ag/AgCl RE, a Pt CE and a polymer coated gold sensor as WE. All polymers were equilibrated by five cyclic voltammetry cycles from -0.3 V to 0.6 V at a scan rate of 20 mV s$^{-1}$. Frequency data and dissipation shifts were collected on 5$^{th}$, 7$^{th}$ and 9$^{th}$ overtone for all polymers on QSotf401, and the data is further analyzed and fitted by D-find software.

Coarse-Grained Molecular Dynamics (CGMD) simulations: The CG model of Savoie and coworkers for mixed ion-electron conductors in aqueous solutions[47,48] was modified to represent the P3CBT polymer. The Savoie model is based on the Martini CG forcefield[49] and incorporates an ellipsoidal CG bead to represent the planar thiophene ring. In this work, the outermost bead of the sidechain was changed to a Martini P3 bead type to mimic the polar COOH functionality of P3CBT. Simulations of a single polymer comprised of 20 P3CBT monomers were performed using the LAMMPS software.[50] The polymer chain was dissolved in an aqueous solution with explicit Cl$^-$ and K$^+$ ions added to mimic 0.1 M salt concentration. Additional Cl$^-$ ions were added to neutralize the degree of backbone charge, which ranged from q = 0.05 to q = 0.2 per monomer to mimic different oxidation states along the doping process. To simulate the high pH deprotonated system, the end groups of the sidechains were changed to Martini bead type Qa, which carries a negative charge. The appropriate number of K$^+$ ions were added in the deprotonated case to maintain neutrality. Full details of the simulation methods are given in the SI.

*Operando* Grazing incidence X-ray fluorescence (GIXRF): *Operando* GIXRF spectra were measured at fluorescence mode using Vortex silicon draft detector (SDD) with an incident energy of 11.6 keV to target potassium K-edge. The X-ray beam was focused to a size of 20 µm (vertical) X 100 µm (horizontal) with an incident angle of 0.12 deg, and the average take-off angle was 8.7 deg above the horizontal plane. Fluorescence was collected at an additional 16 deg behind the normal 90 deg orientation, to avoid interference from the WE (pogo pin). The vertical and horizontal movement of the frit cell were motor-controlled, with a horizontal rocking motion of ±0.4 mm applied throughout the process to minimize beam damage to the film. The XRF detector was positioned 65 to 70 mm away from the sample, and the frit cell was housed in a helium-purged environment, isolated by a 25 µm thin polyethylene sheet, with a bubbler through water to maintain helium separation from the SDD.

Polymer Microstructural Analysis: Grazing incidence wide angle X-ray scattering (GIWAXS) was performed at the Stanford Synchrotron Radiation Lightsource (SSRL) beamline 11-3, using an area detector (Rayonix MAR-225) with an incident energy of 12.7 keV. The sample-to-detector distance was 315.9 mm, calibrated using a LaB$_6$ polycrystalline standard. The incidence angle was set to 0.12°, slightly above the critical angle, to sample the full film depth. All X-ray measurements were conducted in a He chamber to minimize air scattering and prevent beam damage to the samples. Raw data was normalized by detector counts and analyzed using the custom python code.

*Operando* X-ray Scattering and Fluorescence: Grazing incidence X-ray fluorescence (GIXRF) and Grazing incidence Wide-Angle X-ray Scattering (GIWAXS) were performed simultaneously using the Frit cell. The incident X-ray beam was focused to a size of ~20 µm (vertical) X ~100 µm (horizontal) with an incident angle of 0.12 deg with an energy of 11.6 keV. Fluorescence collection was performed using Vortex silicon draft detector (SDD) ~70mm from the sample. The detector was set to collect 9 degrees above the sample plane and normal to the incident beam to avoid interference from the WE (pogo pin). The vertical and horizontal movement of the frit cell were motor-controlled, with a horizontal rocking motion of ±0.4 mm applied during data collection to minimize beam damage to the film. The frit cell was housed in a hydrated, helium-purged environment. The potential control during the *operando* measurement was carried out using a PalmSens4 potentiostat with an Ag/AgCl electrode serving as reference and counter electrode and scan rate of 10 mV s$^{-1}$. A stainless steel frit with 20 µm average pore size was mounted on a sample block to facilitate electrolyte penetration. A silicon frit (MakroPor, thickness 350 µm, pore diameter 8 µm, pore size 12 µm) was purchased from MilliporeSigma. A 5 nm Ti adhesion layer and a 50 nm Au layer were then deposited on the frit surface *via* vapor deposition. The polymer was subsequently float-transferred onto the top of the frit, serving as the working electrode.

*Operando* Raman Spectroscopy: Raman spectra were obtained using a Horiba LabRAM Odyssey confocal Raman microscope with a 633 nm excitation laser source in a backscattering geometry. The spectra of the dry films were measured through a 50× objective, while the spectra of films exposed to the electrolyte were measured using a 5× objective. The laser power was set at 50% power (8.5 mW) for all conditions to avoid photothermal effects and sample degradation. For all measurements, thin films of target polymers were spray cast onto ITO/glass substrates (8–12 Ω sq$^{-1}$ Delta Technologies), then acidified as per above and used as working electrodes. Ag/AgCl pellet ($D$ = 2 mm × $H$ = 2 mm, Warner Instruments) and Pt wire were used as RE and CE, respectively. A Gamry interface 1000 Potentiostat was used to perform in situ doping and dedoping experiments. All measurements were taken in an AIST-NT EC001 electrochemical cell filled with a 0.1 M NaCl (aq) electrolyte. The spectral region scanned in this investigation ranged between 600 and 1799 cm$^{-1}$. A silicon wafer was used for the calibration process, and all spectra were collected using LabSpec 6 software and further deconvoluted through PeakFit version 4.12 software.

*Operando* Transient Absorption Spectroscopy: *Operando* transient absorption spectroscopy (TAS) was performed with an Ultrafast Systems Helios spectrometer. One hundred fifty femtosecond (fs) pulses of 800 nm laser light were generated with a Coherent Libra amplified Ti:sapphire system at 1.2 W and 1 kHz repetition rate. Experiments were conducted using a 850 nm pump that was attenuated to 1.0 mW, and a probe generated using a sapphire crystal yielding an optical window of 800 nm - 1600 nm. The transient absorption spectra were measured over a 200 ps window. For each scan, 100 time points were recorded with exponential spacing, and each sample was subjected to three scans that were averaged to create the complete dataset. A reference point was used to align reflectance of the sample in the holder to maintain consistency in the angle of incident for the pump beam for each experiment. Select scans for each sample were compared before analysis to ensure that no degradation of signal had occurred during the experiment and ensure data consistency. The TAS signal intensity is typically presented as a change in absorption (ΔA), which can broadly be understood as the difference between the excited-state absorption and the ground-state absorption. Data preparation, including chirp correction were performed using the global analysis software Surface Xplorer provided free of charge by Ultrafast systems. A Gamry interface 1000 Potentiostat was used to perform *in situ* doping and dedoping experiments. All measurements were taken in quartz cuvettes and filled with electrolyte.

*Operando* Atomic Force Microscopy (AFM) and Force Measurements: AFM measurements were performed *in situ* in the electrolytes listed using an Oxford Instruments/Asylum Research Cypher-ES. All cantilevers were calibrated on a sapphire reference sample in the same electrolyte to correct for spring constant and optical lever sensitivity, using a gold-coated contact mode cantilever (BudgetSensors ContGB-G, k ~0.2 N/m, ambient $f_0$~13 kHz) for force mapping and Cr/Pt-coated cantilevers (BudgetSensors, Multi75E-G k ~2 N/m, ambient $f_0$ ~75 kHz). For height measurements, data were taken in AC mode using using a razor blade scratch on ITO as a reference. The AFM chamber was purged with nitrogen prior to being sealed. The bias voltage in all data was applied to the cantilever. Data were analyzed using Igor Pro.

*Operando* ultraviolet-visible-near-infrared (UV-Vis-NIR) spectroelectrochemistry: measurements were performed using the same setup as the CV experiment, with a three-electrode configuration in aqueous solution under a nitrogen atmosphere. The WE and RE were polymer coated ITO and Ag/AgCl, respectively, consistent with the CV experiment described above, with a platinum wire (Gamry) used as the CE. The experiment was conducted using an Agilent Cary 5000 spectrophotometer with quartz cuvettes having a path length of 1 cm. The polymer thin film was biased under potentiostatic conditions, controlled by Gamry Interface 1010B potentiostat. The film spectra were recorded once WE current reached a steady state which typically required 30 to 60 seconds for each potential step.

OECT fabrication and characterization: Interdigitated organic electrochemical transistor (iOECT) substrates were purchased from MicruX technologies (Model ED-IDE1-Au, total area = 38.5 mm$^2$, 10 µm electrode length, 10 µm electrode gap, gold thickness = 150 nm, active area = 9.6 mm$^2$, average width =

2.75 mm).and devices were fabricated using prior published methods.[29] A cylinder-shaped Ag/AgCl pellet (Warner Instruments) was used as a gate electrode and immersed in a 0.1 M aqueous NaCl solution confined in a PDMS well. Transfer and output characteristics were measured with an Agilent B1500A semiconductor analyzer. All transistor characteristics were collected by using Keysight Easy Expert software with a custom *I*/*V* sweep configuration and were tested under ambient conditions. Data were collected for at least five separate devices to ensure reproducibility. Output curves were collected with a $V_d$ step size −0.1 mV, and a $V_g$ step size of −0.1 V. Transfer curves were collected using a $V_d$ = −0.6 V with a $V_g$ step size of −0.1 mV.

## Acknowledgements:

The authors appreciate support from the National Science Foundation (NSF) Grant No. 2408881. In addition, ER, TEG, ERY, MS, ZS, SQ, and RL acknowledge partial support from Lehigh University, funds associated with the Carl Robert Anderson Chair in Chemical Engineering, and access to the Lehigh University Institute for Functional Materials and Devices (I-FMD) Materials Characterization Facility (MCF) and Integrated Nanofabrication and Cleanroom Facility (INCF); and YZ appreciates partial support from the Georgia Institute of Technology. Use of the Stanford Synchrotron Radiation Lightsource, SLAC National Accelerator Laboratory, is supported by the U.S. Department of Energy, Office of Science, Office of Basic Energy Sciences under Contract No. DE-AC02-76SF00515. ERY thanks the NSF Major Research Instrumentation program for funds that established the multi-user laser facility for transient absorption (CHE-1428633) and ERY and RL appreciate NSF support for funding the transient absorption spectroscopy research efforts of this manuscript (CHE-2313290). DSG and RG acknowledge NSF for supporting the AFM work (DMR-2309577). Part of this work was conducted at the University of Washington Molecular Analysis Facility, a National Nanotechnology Coordinated Infrastructure site, which is supported in part by the National Science Foundation (NNCI-1542101), the University of Washington, the Molecular Engineering & Sciences Institute, the Clean Energy Institute, and the National Institutes of Health. TEG thanks Prof. Brett Savoie for sharing the implementation of the customized LAMMPS ellipsoidal bond and angle potentials. The authors thank Dr. Nathan Wittenberg and Dane Santa for assistance with QCM-D measurement. The authors also thank Dilara Meli, Ruiheng Wu and Joseph Strzalka for highly fruitful discussions and assistance with electrochemical frit cell setup. The authors also thank Dr. Tom van der Pol for the discussion of the transient mass response.

Supporting information:

# pH Regulates Ion Dynamics in Carboxylated Mixed Conductors

Zeyuan Sun,[1] Mengting Sun,[1] Rajiv Giridharagopal,[2] Robert C. Hamburger,[3] Siyu Qin,[1] Haoxuan Li,[1] Mitchell C. Hausback,[1] Yulong Zheng,[4] Bohyeon Kim,[1] Heng Tan,[5] Thomas E. Gartner III,[1*] Elizabeth R. Young,[3*] Christopher J Takacs,[6*] David S. Ginger,[2*] Elsa Reichmanis[1*]

[1]Department of Chemical and Biomolecular Engineering, Lehigh University, Bethlehem, PA 18015, United States

[2]Department of Chemistry, University of Washington, Seattle, WA 98195, United States

[3]Department of Chemistry, Lehigh University, Bethlehem, PA 18015, United States

[4]School of Chemistry and Biochemistry, Georgia Institute of Technology, Atlanta, GA 30332, United States

[5]Department of Computer Science and Engineering, Lehigh University, Bethlehem, PA 18015, United States

[6]Stanford Synchrotron Radiation Lightsource SLAC National Accelerator Laboratory, Menlo Park, CA 94025, United States

# Table of Contents

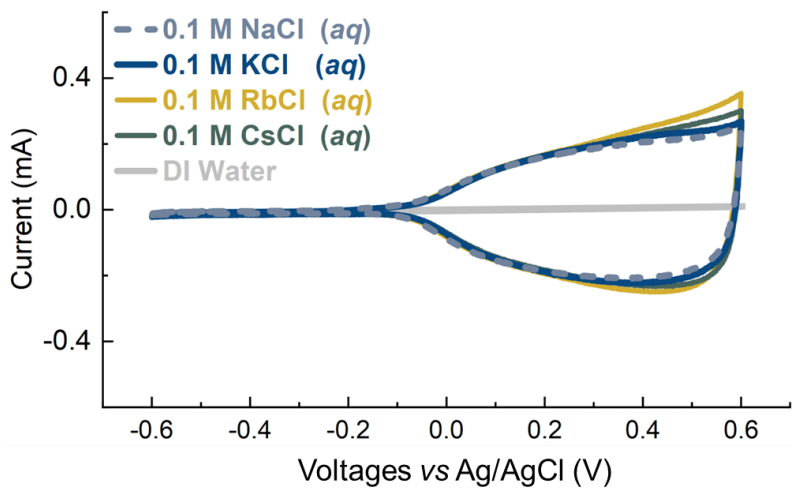

**Figure S1**: **Cation-independent behavior**. Cyclic voltammetry (CV) was performed on P3CBT-P in chloride-based aqueous electrolytes containing different monovalent cations, under conditions of neutral pH and comparable thickness. The voltammogram reveals a largely cation-independent electrochemical response, with sodium and potassium electrolyte exhibiting negligible differences in performance – supporting their use as valid reference for $PF_6$ -based system. Slight increases in peak current were observed with rubidium and cesium chlorides, but all chloride salts showed similar turn-on voltage. In contrast measurements in deionized water, used as a background control, showed no detectable conductivity.

### *Concentration calculation based on pH titration methods:*

The pH of 100 ml of 0.1 M KCl electrolyte is regulated through titrating concentrated HCl solution, where the pH is related to the concentration of protons by the equation:

Molarity of concentrated HCl, where the 37% w/w HCl solution, with 37 g per 100 g solution, with a density of 1.19 g/ml, and a molar mass of 36.46 g/mol

Therefore, the molarity of HCl is:

$$HCl = \frac{37g \div 36.46 g/mol}{100\ g\ \div 1.19\ g/ml} = \frac{1.015\ mol}{0.08403\ L} = 12.08\ M$$

**Moles of HCl required to reach pH 2.5:**

$$[H^+] = 10^{-2.5} = 3.16 \times 10^{-3} mol/L$$

**To reach this proton concentration in ~100 ml of solution:**

$$mol\ HCl\ needed = 3.16\ \times 10^{-3} mol/L\ \times 0.1\ L = 3.16 \times 10^{-4} mol$$

$$Volume\ of\ 12.08\ M\ HCl\ added = \frac{3.16\ \times 10^{-4} mol}{12.08\ mol/L} = 26.2\ \mu L$$

**Total moles of Cl⁻ in the final solution:**

$$[Cl^-] = \frac{0.010\ mol + 0.000316\ mol}{100\ ml + 26.2\ \mu L} = 0.1031\ M$$

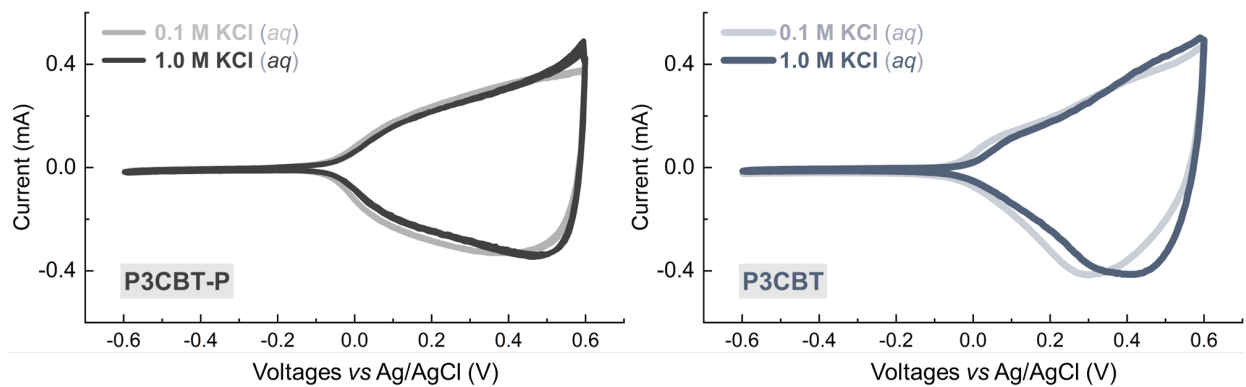

**Figure S2**: **Concentration-independent behavior**. Cyclic voltammetry (CV) was performed on P3CBT-P and P3CBT in chloride-based aqueous electrolytes of varying concentration (0.1 – 1 M), with neutral pH and comparable film across samples. The voltammogram show that changes in potassium or chloride ion within this range do not significantly affect the electrochemical response, indicating that ion concentration does not play a significant role in the pH-dependent behavior throughout this study, and the voltage offset shown in main figure 1 is mainly due to the pH induced dissociation change.

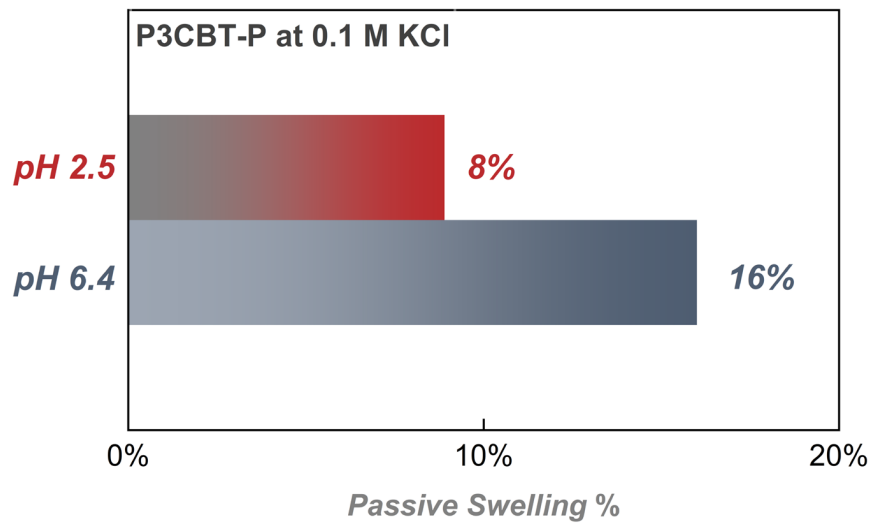

**Figure S3**: **Experimental hydrophilicity study**. Comparison of passive swelling percentage comparison for P3CBT-P in 0.1 M acidic KCl (red) and neutral KCl (blue). Combined with GIXRF data, this result indicates reduced cation uptake under acidic conditions. This suggests that passive swelling is primarily influenced by cation-polymer interactions and the intrinsic hydrophilicity (degree of dissociation) of the polymer.

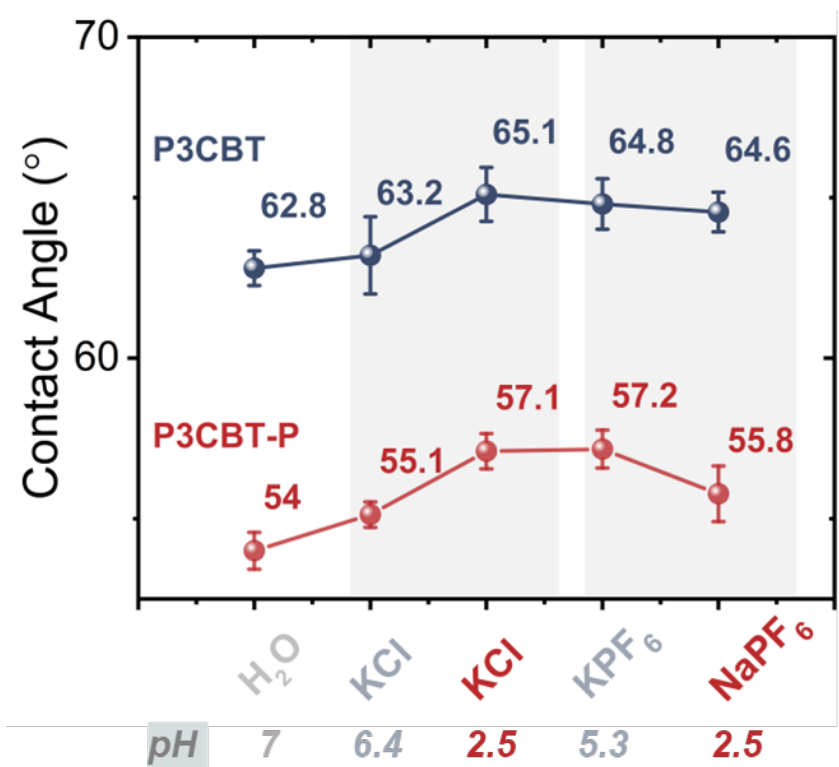

**Figure S4**: Contact angle measurement of thin films for P3CBT-P (red) P3CBT (blue), and P3EPT (light brown) average from five samples in $H_2O$, 0.1 M KCl at neutral pH, 0.1 M KCl at acidic pH, and 0.1 M $KPF_6$ at neutral pH and $NaPF_6$ at acidic pH

**Discussion 1: Coarse-Grained Model**

We created a model for P3CBT based on the model developed by Savoie and coworkers for P3HT.[1] The Savoie model is itself based on the MARTINI coarse-grained (CG) force field[2] and added ellipsoidal species to capture the anisotropic nature of thiophene rings.[3] In this work, we adjusted the P3HT model to represent P3CBT by changing the side chain species. The P3CBT system was represented using six main components: backbone, non-polar side chain, polar side chain, anion, cation, and solvent. Each polymer monomer was mapped to three CG beads: one anisotropic ellipsoidal backbone bead (BB), one spherical non-polar side chain bead (NS), and one spherical polar side chain bead (PS).

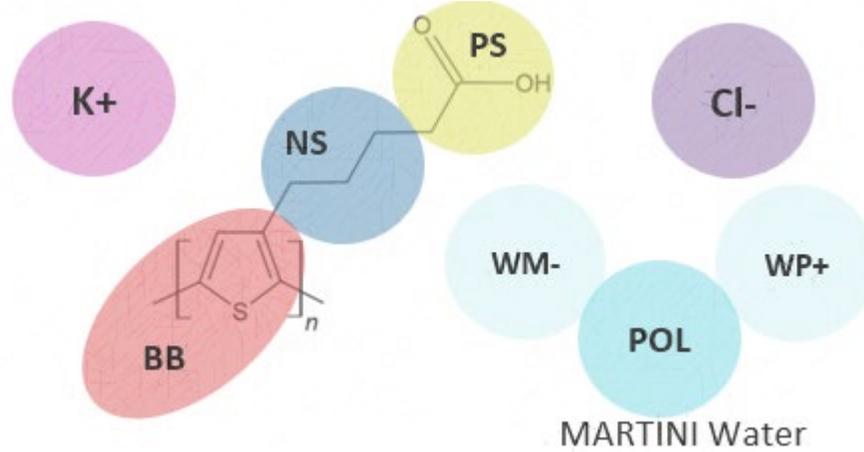

*Figure shows Coarse-grained representation of the P3CBT system*

Non-bonded interactions between spherical CG beads were computed using the Lennard-Jones (LJ) and Coulomb potentials with shifting cutoff functions $S_{LJ}(r)$ and $S_C(r)$ shown in **Equations 1 to 3**.[4] The LAMMPS pair style "lj/gromacs/coul/gromacs" was employed.[5]

$$U_{12}(r) = 4\varepsilon\left[\left(\frac{\sigma}{r}\right)^{12} - \left(\frac{\sigma}{r}\right)^6\right] + S_{LJ}(r) \tag{1}$$

$$U_{12}(r) = \frac{Cq_1q_2}{\epsilon r} + S_C(r) \tag{2}$$

Where $U_{12}$ is the potential between particles 1 and 2, $r$ is the center-to-center distance between particles, $\sigma$ is the distance at which the interaction potential equals zero, $\varepsilon$ is the depth of the energy well, C is a coefficient computed by LAMMPS, and $q_1$, $q_2$ are the charges of particles 1 and 2. The LJ potential used an inner cut-off of 9 Å and an outer cut-off of 12 Å. For the Coulomb potential, only an outer cut-off of 12 Å was applied, with the inner cut-off set to 0 Å.

Backbone-backbone interactions between ellipsoidal beads were treated with the Gay-Berne (GB) potential to capture the anisotropic interactions with semi-axis lengths $\sigma_x$, $\sigma_y$, and $\sigma_z$.

$$U_{12}(\omega_1, \omega_2, r_{12}) = 4\varepsilon_0 \varepsilon_{12}^v(\omega_1, \omega_2)\varepsilon_{12}'^\mu(\omega_1, \omega_2, \hat{r}_{12})$$

$$\times \left\{\left[\frac{\sigma_c}{r_{12} - \sigma_{12}(\omega_1, \omega_2, \hat{r}_{12}) + \sigma_c}\right]^{12} - \left[\frac{\sigma_c}{r_{12} - \sigma_{12}(\omega_1, \omega_2, \hat{r}_{12}) + \sigma_c}\right]^6\right\} \tag{3}$$

$\omega_1$ and $\omega_2$ are the Euler angles of the particles, $r_{12}$ is the displacement vector, $\varepsilon_{12}$ and $\varepsilon'_{12}$ are energy prefactors, $\sigma_c$ is the minimum contact distance, and $\varepsilon_0$ is the well depth. The exponents v and μ are empirical coefficients that are set to one. A 12 Å cutoff was applied. Directly bonded neighbors within a given molecule were excluded from the nonbonded interaction calculation.

Bonded potentials involving backbone beads were computed using custom LAMMPS classes developed by the Savoie group[3]: "elel" (ellipsoid-ellipsoid, backbone-backbone interaction) and "elsp" (ellipsoid-sphere, backbone-side chain interaction). All other bonded interactions used standard LAMMPS potentials, namely harmonic bonds, cosine-squared angles, and OPLS dihedrals. The full set of bonded parameters is provided in **Table S1**.

**Table S1**: Bonded interaction parameters. Units of all parameters are consistent with LAMMPS "real" units (kcal/mol for $k$, Å for $r$). Angle units are in degrees.

| Bonds | | Angles | |
|---|---|---|---|
| BB-BB | $k = 25.0, r_0 = 4.3$ | BB-BB-BB | $k_a = 1.0, \theta_0 = 160$ |
| BB-NS | $k = 1.5, r_0 = 4.7$ | BB-BB-NS | $k_a = 3.0, \theta_0 = 90$ |
| NS-PS | $k = 1.5, r_0 = 4.6$ | BB-NS-PS | $k_a = 3.0, \theta_0 = 180$ |
| POL-WP, POL-WM | $k = 1.0, r_0 = 1.4$ | WM-POL-WP | $k_a = 0.5019, \theta_0 = 0$ |
| Dihedrals | | Parameters for Custom Potentials | |
| BB-BB-BB-BB | $K_1 = 0.5, K_2 = 0.6$ $K_3 = 0.0, K_4 = 0.0$ | BB-BB, $d$ | $K_1 = 0.0, K_2 = 2.0$ $K_3 = 0.0, K_4 = 0.0$ |
| NS-BB-BB-NS | $K_1 = 3.0, K_2 = 0.0$ $K_3 = 0.0, K_4 = 0.0$ | BB-BB, $a_1$ | $k_a = 7.5, r_0 = 90$ |
| | | BB-BB, $a_2$ | $k_a = 7.5, r_0 = 90$ |
| | | BB-NS, $a$ | $k_a = 3.0, r_0 = 90$ |

The size, mass, and charge of each coarse-grained bead were specified through the LAMMPS atom style. A custom atom style, *hybrid full ellipsoid*, was used to define ellipsoidal backbone beads, particle charges, and molecular connectivity. To enable these features, the MOLECULE, ASPHERE, and EXTRA-PAIR packages were required during LAMMPS compilation. The bead mass is not assigned directly in LAMMPS but instead calculated from bead volume and density shown in **Equation 4**:

$$Mass = Volume \times Density = \frac{4}{3}\pi R_x R_y R_z \times Density \quad (4)$$

Where $R_x$, $R_y$, and $R_z$ are the bead raddi in the x, y, and z directions. The diameter and density of each bead are defined within the LAMMPS input script.

**Parameterization**

Nonbonded interactions were parameterized using the MARTINI CG force field, originally developed for lipid and biomolecular systems but now widely applied to conjugated polymers.[2] The MARTINI framework provides a transferable library of bead types calibrated against experimental partitioning free energies, while maintaining computational efficiency by grouping about four heavy atoms into a single coarse-grained site. This mapping preserves essential chemical properties such as polarity, charge, and hydrogen-bonding capability, while allowing simulations to reach the larger system sizes and longer timescales. Within the MARTINI force field, bead types are categorized according to polarity: polar (P), nonpolar (N), apolar (C), or charged (Q). Each category is further refined by numerical indices (1–5, from least to most polar) or hydrogen-bonding properties: "a" for acceptor, "d" for donor, "da" for both, and "0" for none. In addition, the force field defines a set of "small" beads, designated with the prefix *S*, which represent a reduced effective volume compared to standard beads and are often used to describe sterically restricted groups such as short sidechains. The MARTINI bead assignments for each component in our system are listed in **Table S2**.

**Table S2**: MARTINI bead assignments

| Component | MARTINI Bead |
|---|---|
| Backbone (BB) | C4 |
| Non-polar Sidechain (NS) | SC3 |
| Polar Sidechain (PS) | P3 |
| Anion (Q) | Qa |
| Cation (Q) | Q0 |

Each thiophene unit in P3CBT was represented as an ellipsoidal bead. The in-plane dimensions of the ellipsoid (x and y) were set to 5 Å, corresponding to the approximate diameter of a thiophene ring, while the out-of-plane dimension (z) was set to 3 Å, consistent with the average π–π stacking distance.[6,7] This anisotropic representation allows the coarse-grained model to reproduce direction-dependent interactions critical for backbone ordering. All other components of the system were modeled as finite-sized spherical beads. For spherical beads, the diameter was chosen to match the self-interaction zero-energy distance ($\sigma$) from the LJ potential model. In this system, spherical beads were assigned to a diameter of 4.7 Å, except for the non-polar sidechain bead, which was slightly smaller at 4.3 Å.

Cations and anions in the model served both as explicit KCl salt and as compensating charges for the polymer. Potassium ions ($K^+$) were assigned a charge of +1e, and chloride ions ($Cl^-$) a charge of –1e. Additional anionic or cationic beads (–1e or +1e) were introduced as needed to maintain charge neutrality, either balancing the positively charged backbone or the negatively charged polar sidechains in deprotonated systems. Consistent with previous coarse-grained models of OMIECs, the oxidative charge on the polymer backbone was assumed to be uniformly distributed across all monomers.[8]

The solvent was represented using the MARTINI polarizable water model.[9] In this model, three coarse-grained beads correspond to four real water molecules, enabling explicit treatment of orientational polarizability. The WP and WM beads carry charges of +0.46e and –0.46e, respectively, and a global dielectric constant of 2.5 was applied. This water representation allows for a more realistic description of hydration and ion solvation compared to nonpolarizable models. For MARTINI polarizable water, intramolecular WP/WM interactions were excluded to preserve rotational polarizability. The parameters for all non-bonded interactions, including solvent, are provided in **Table S3**.

**Table S3**: Non-bonded interaction parameters. In LAMMPS "real" units (kcal/mol for $\varepsilon$, Å for $\sigma$).

| | BB | NS | PS | POL | Qa | Q0 |
|---|---|---|---|---|---|---|
| BB | $\sigma_x = \sigma_y = 5.0, \sigma_z = 3.0$<br>$\varepsilon_x = \varepsilon_y = 0.25, \varepsilon_z = 1.2$<br>$\varepsilon_0 = 1.0, \sigma_c = 3.0$ | $\varepsilon = 0.84$<br>$\sigma = 4.7$ | $\varepsilon = 0.74$<br>$\sigma = 4.7$ | $\varepsilon = 0.61$<br>$\sigma = 4.7$ | $\varepsilon = 0.74$<br>$\sigma = 4.7$ | $\varepsilon = 0.74$<br>$\sigma = 4.7$ |
| NS | | $\varepsilon = 0.63$<br>$\sigma = 4.3$ | $\varepsilon = 0.74$<br>$\sigma = 4.7$ | $\varepsilon = 0.61$<br>$\sigma = 4.7$ | $\varepsilon = 0.65$<br>$\sigma = 4.7$ | $\varepsilon = 0.65$<br>$\sigma = 4.7$ |
| PS | | | $\varepsilon = 1.19$<br>$\sigma = 4.7$ | $\varepsilon = 1.14$<br>$\sigma = 4.7$ | $\varepsilon = 1.34$<br>$\sigma = 4.7$ | $\varepsilon = 1.34$<br>$\sigma = 4.7$ |
| POL | | | | $\varepsilon = 0.96$<br>$\sigma = 4.7$ | $\varepsilon = 1.19$<br>$\sigma = 4.7$ | $\varepsilon = 1.07$<br>$\sigma = 4.7$ |
| Qa | | | | | $\varepsilon = 0.84$<br>$\sigma = 4.7$ | $\varepsilon = 0.55$<br>$\sigma = 4.7$ |
| Q0 | | | | | | $\varepsilon = 0.84$<br>$\sigma = 4.7$ |

The effect of pH was modeled by varying the protonation state of the carboxylic acid end group on the sidechains. The only difference between the acidic and neutral conditions was the MARTINI bead representation of the carboxylic acid group. The Henderson-Hasselbalch Equation (**Equations 5 and 6**) was used to calculate the percentage of deprotonation for a given pH. Assuming a pKa of 4.5 for the carboxylic acid, experimental conditions at pH 2.5 and 6.4 correspond to calculated deprotonation percentages of ~1% and ~99%, respectively. Accordingly, the acidic condition (pH 2.5) was modeled as fully protonated, where the sidechain end group was represented by a neutral P3 bead. The neutral condition (pH 6.4) was modeled as fully deprotonated, where the sidechain end group was represented by a negatively charged Qa bead corresponding to COO⁻.

$$pH = pKa + \log_{10}\left(\frac{[A^-]}{[HA]}\right) \quad (5)$$

$$Deprotonated\ \% = \frac{[A^-]}{[A^-] + [HA]} \times 100\% = \frac{10^{pH-pKa}}{10^{pH-pKa} + 1} \times 100\% \quad (6)$$

**Molecular Dynamics**

All molecular dynamics simulations were performed using LAMMPS[10] with a 10 fs timestep and Velocity–Verlet integration under periodic boundary conditions. Systems were initialized in a cubic simulation box with a 4050 Å box side length using Moltemplate,[11] which placed molecules at random locations chosen from a diffuse grid. A 150 Å lattice spacing along all three axes were used to ensure components were non-overlapping. The bead velocities are initialized using a randomly seeded uniform distribution to give a system temperature of 300K. The system was equilibrated through a multi-stage protocol. First, an NVE relaxation was performed for 10 ps with bead displacements limited to 0.1 Å per timestep. This was followed by a 1 ns NVT simulation in which the simulation box was linearly deformed until the system density reached 1.0 g/cm³. After this step, beads velocities were rescaled to 300 K using another randomly seeded uniform distribution to prevent the buildup of translational or rotational momentum. A second short NVE relaxation of 50 ps with displacement restraints was then carried out, after which the system was equilibrated in the NPT ensemble at 300 K and 1 atm for 5 ns. To stabilize the density, an additional NVT run of 5 ns was performed at 300K using the average density from the NPT step. Finally, production simulations were carried out in the NPT ensemble at 300 K for 50 ns. Trajectory data in the production run was recorded every 10,000 timesteps, corresponding to an interval of 100 ps. This resulted in 500 frames for radial distribution function, g(r) analysis. Throughout equilibration and production, the linear and angular momentum of the system were zeroed every 10 ps.

Each simulation contained 1 polymer chain of 20 monomers each, along with 65 $K^+$ cations and $Cl^-$ anions, and 9000 MARTINI water molecules (corresponding to 36000 real water molecules). The ion concentration was chosen to represent a background electrolyte composition of 0.1 M. Additional ion content was adjusted according to backbone charge and protonation state to maintain overall system charge neutrality. Three independent replicates were performed for each of the four states considered in this work: two backbone charge states (q = 0.05e and q = 0.2e) and two deprotonation states (fully protonated and fully deprotonated).

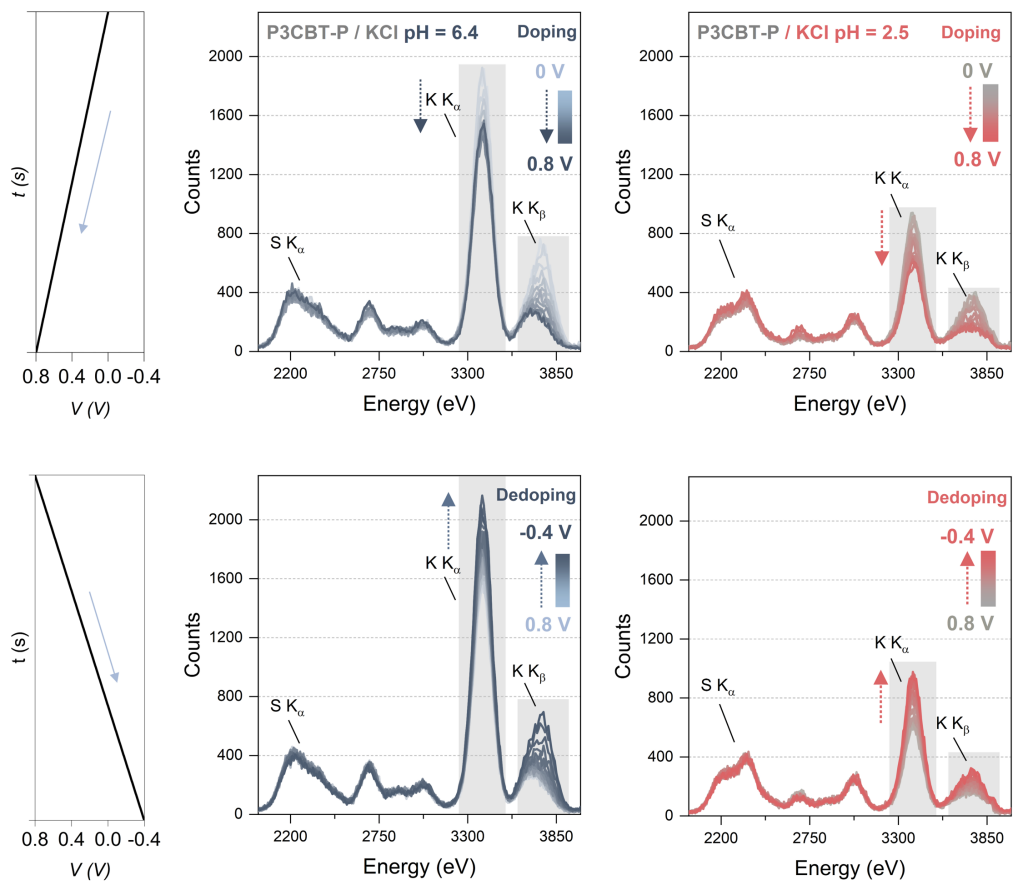

**Figure S5**: *Operando* **GIXRF during (de)doping**. Raw data for P3CBT-P in 0.1 M KCl at pH 6.4 (blue), and pH 2.5 (red), collected during doping (0 V to 0.8 V, top row), and dedoping (0.8 V to -0.4 V, bottom row) at a rate of 10 mV s$^{-1}$.

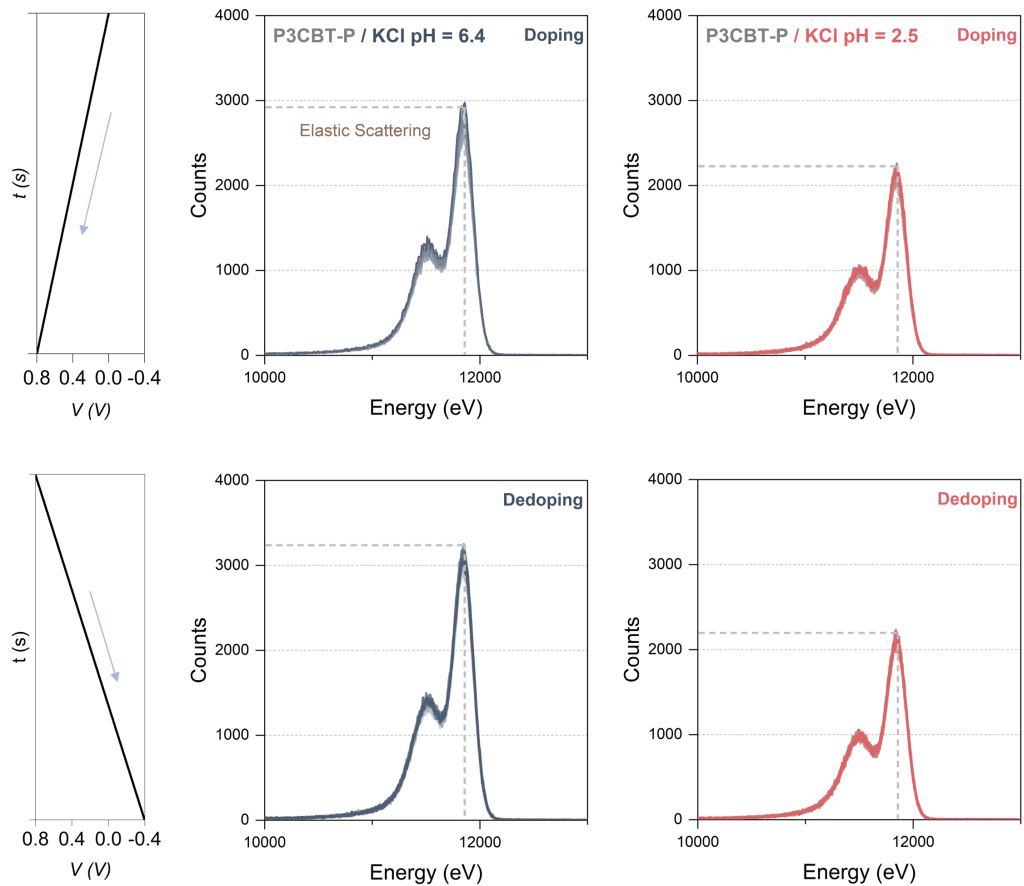

**Figure S6**: Elastic scattering peak from the *operando* GIXRF data in **Fig. S5**, used for beam energy calibration and normalization purpose to enable qualitative comparison.

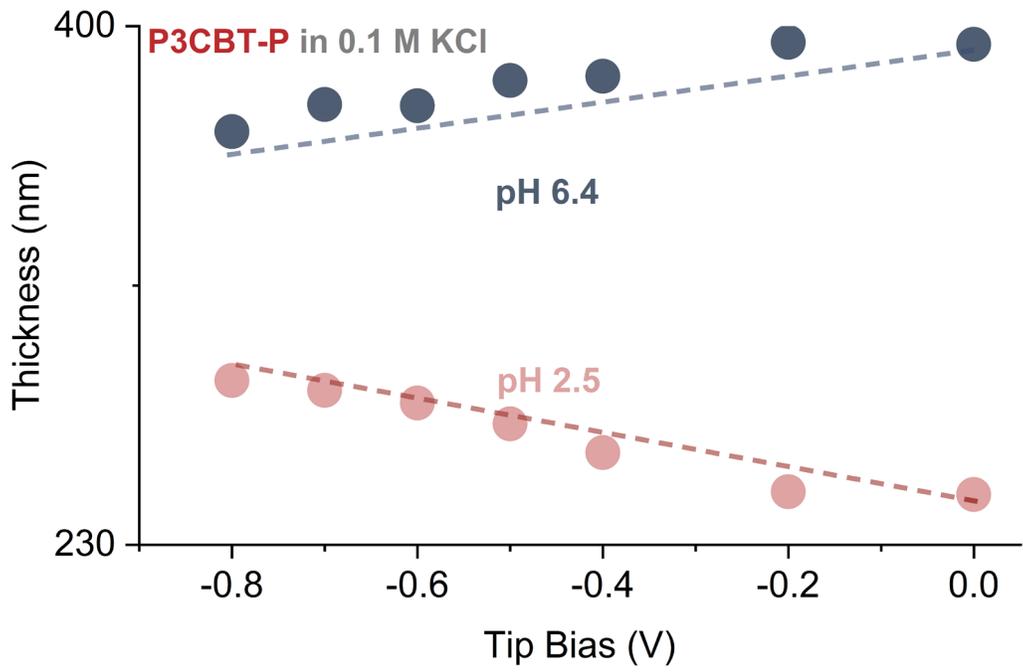

**Figure S7**: Absolute thickness changes in in-situ AFM for P3CBT-P under neutral and acidic pH conditions.

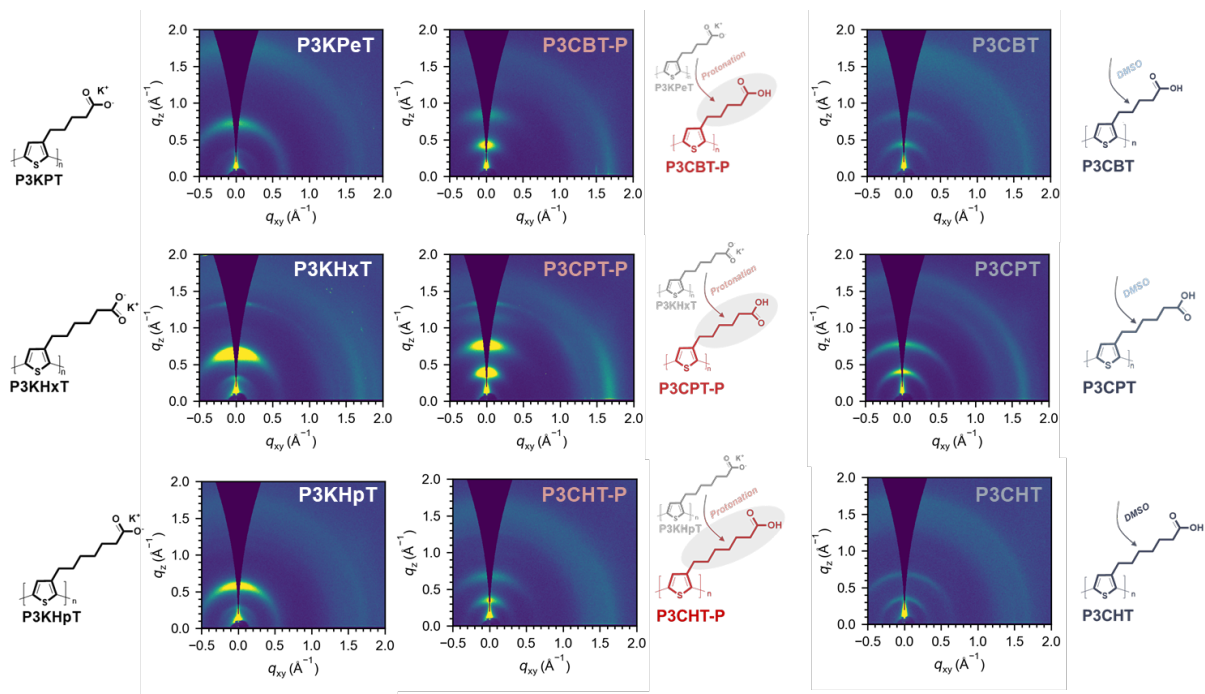

**Figure S8: 2D GIWAXS patterns of carboxylated polythiophenes.** The left column shows the **COOK** series (water soluble carboxylate form): P3KPeT, P3KHxT, and P3KHpT from top to bottom. The middle column presents the **protonated COOH** series (carboxylic acid form by acidification): P3CBT-P, P3CPT-P, and P3CHT-P. The right column displays the **pristine COOH** series (carboxylic acid in DMSO): P3CBT, P3CPT, and P3CHT from top to bottom.

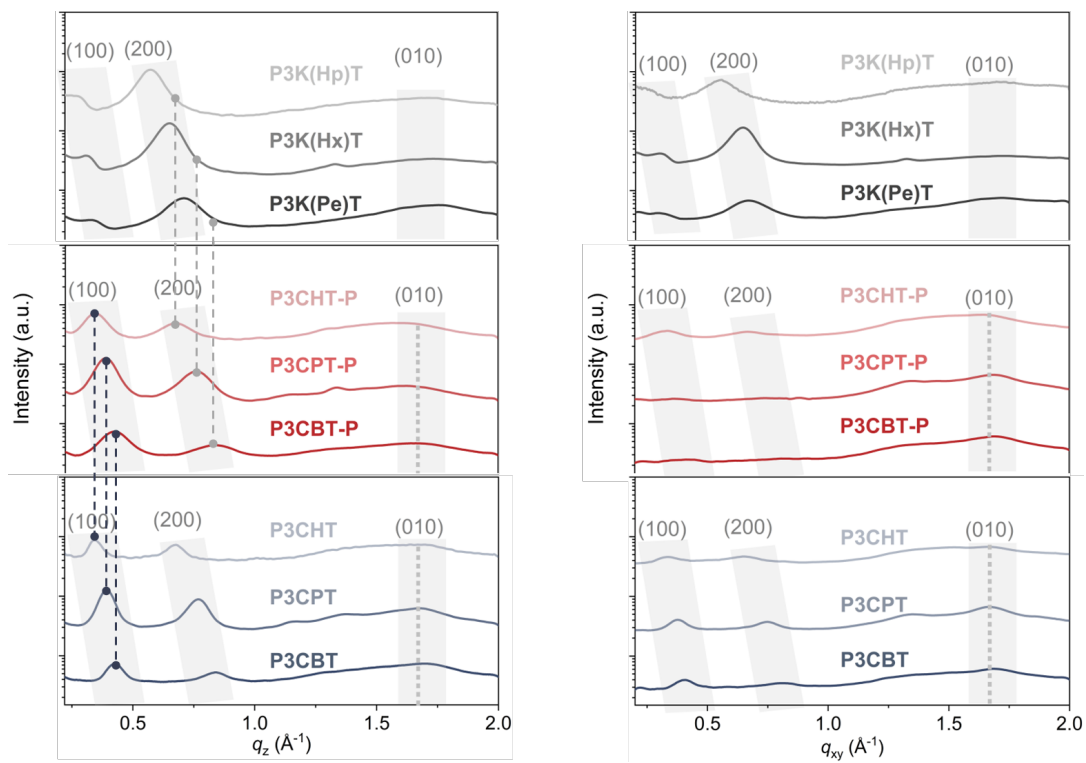

**Figure S9:** the corresponding 2D line cut profiles for all the series respectively.

These Figure **S8** and **S9** results indicate that:

1) Increasing the side chain spacer significantly expands the lamellar spacing, while π–π stacking remains largely unaffected across all polymers.

2) As previously shown,[12] the K-to-H conversion significantly influences molecular arrangement and packing. The resulting change in lamellar spacing is expected due to volumetric effects from ion exchange. With longer side chain spacers, the protonated COOH polymers become more isotropic, with P3CHT-P exhibiting the highest isotropy among the series. In contrast, the DMSO-treated COOH series displays consistently isotropic microstructures across all variants, differing from the protonated counterparts.

3) The peak positions remain unaffected by processing conditions, with all COOH polymers exhibiting identical lamellar and π–π stacking peaks.

4) Microstructurally, the longest side-chain polymer serves as a representative for both series due to its structural similarity and comparable/beneficial crystallinity features observed in X-ray scattering experiments.

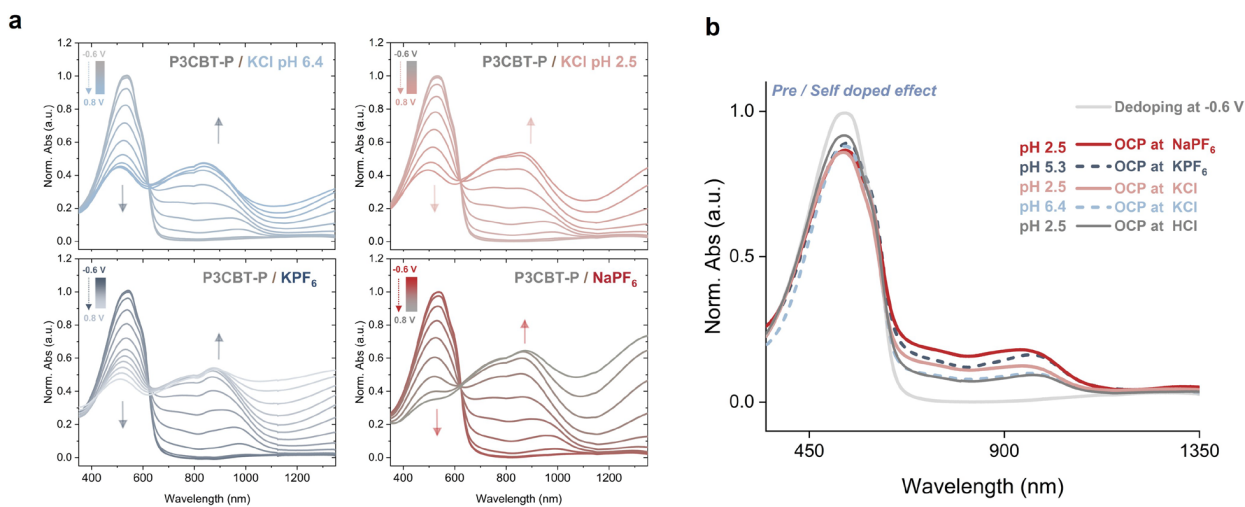

**Figure S10**: *Operando* **spectroelectrochemistry and pre- / self-doped evidence.** (a) Spectroelectrochemical response of P3CBT-P in pH-neutral electrolytes (0.1 M KCl and 0.1 M KPF$_6$, left) and pH-acidic electrolytes (KCl and NaPF$_6$, right). (b) UV–Vis spectra at open-circuit potential show polaronic absorption bands (800–1000 nm), indicating self-doping behavior. The intensity and position of these bands are both ion- and pH-dependent, with acidic conditions exhibiting a higher pre-doped state likely due to reduced cation–polymer interactions.

## Transient absorption spectroscopy (TAS).

For samples of P3CBT-P under either neutral or acidic pH conditions, the results of the chirp correction to TAS data (2-panel figures) are shown below. A comparison of traces at 1000 nm are also presented to demonstrate the rate of the decay at both OCP and 0.8 V samples under neutral and acidic conditions. The TAS signal intensity is typically presented as a change in absorption (ΔA), which can broadly be understood as the difference between the excited-state absorption and the ground-state absorption. In cases when ΔA is relatively small, the change in absorption is denoted with a milli suffix (mΔA, in which mΔA = $10^{-3}$ ΔA units).

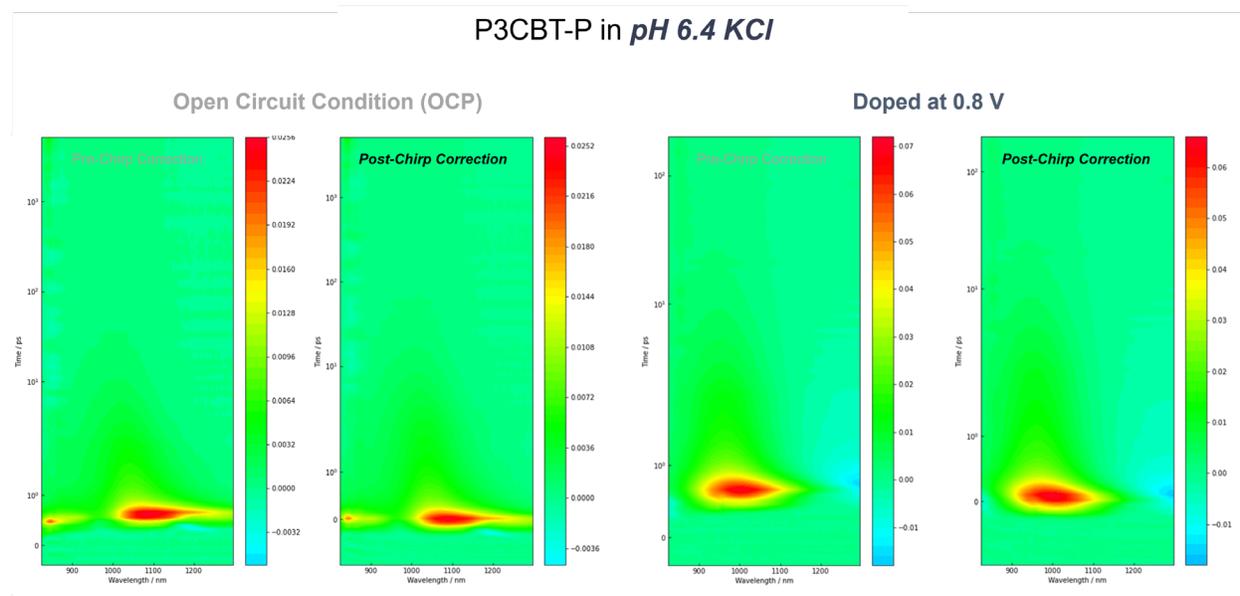

**Figure S11:** TAS raw data for carboxylated polythiophene under pH neutral condition. Raw data surface of the P3CBT-P sample in neutral pH 0.1 M KCl, shown at open-circuit potential (OCP, left) and under doping conditions (right), before and after chirp correction.

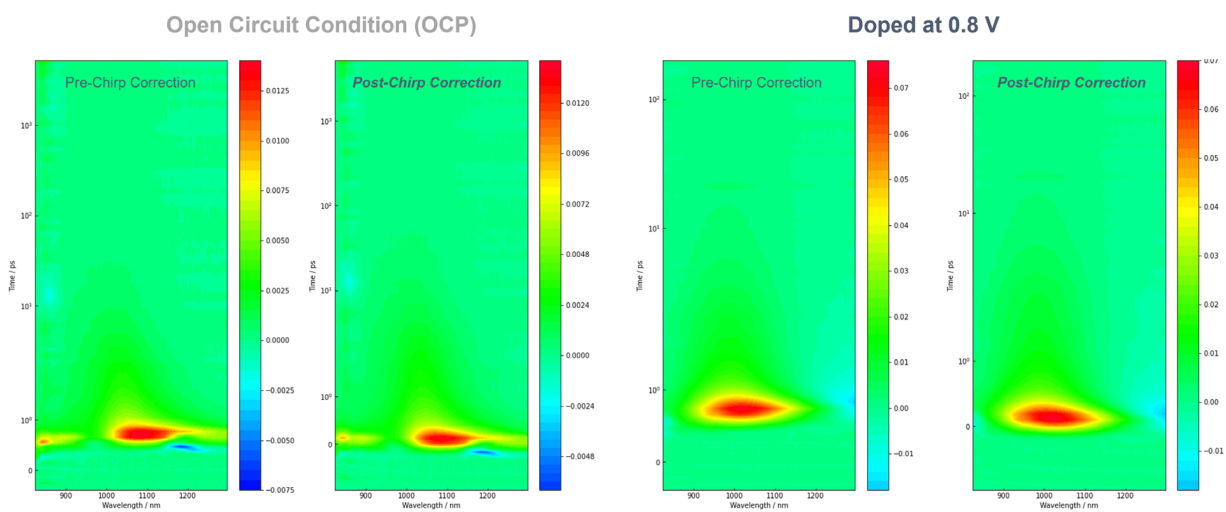

**Figure S12:** TAS raw data for carboxylated polythiophene under pH acidic condition. Raw data surface of the P3CBT-P sample in acidic pH 0.1 M KCl, shown at open-circuit potential (OCP, left) and under doping conditions (right), before and after chirp correction.

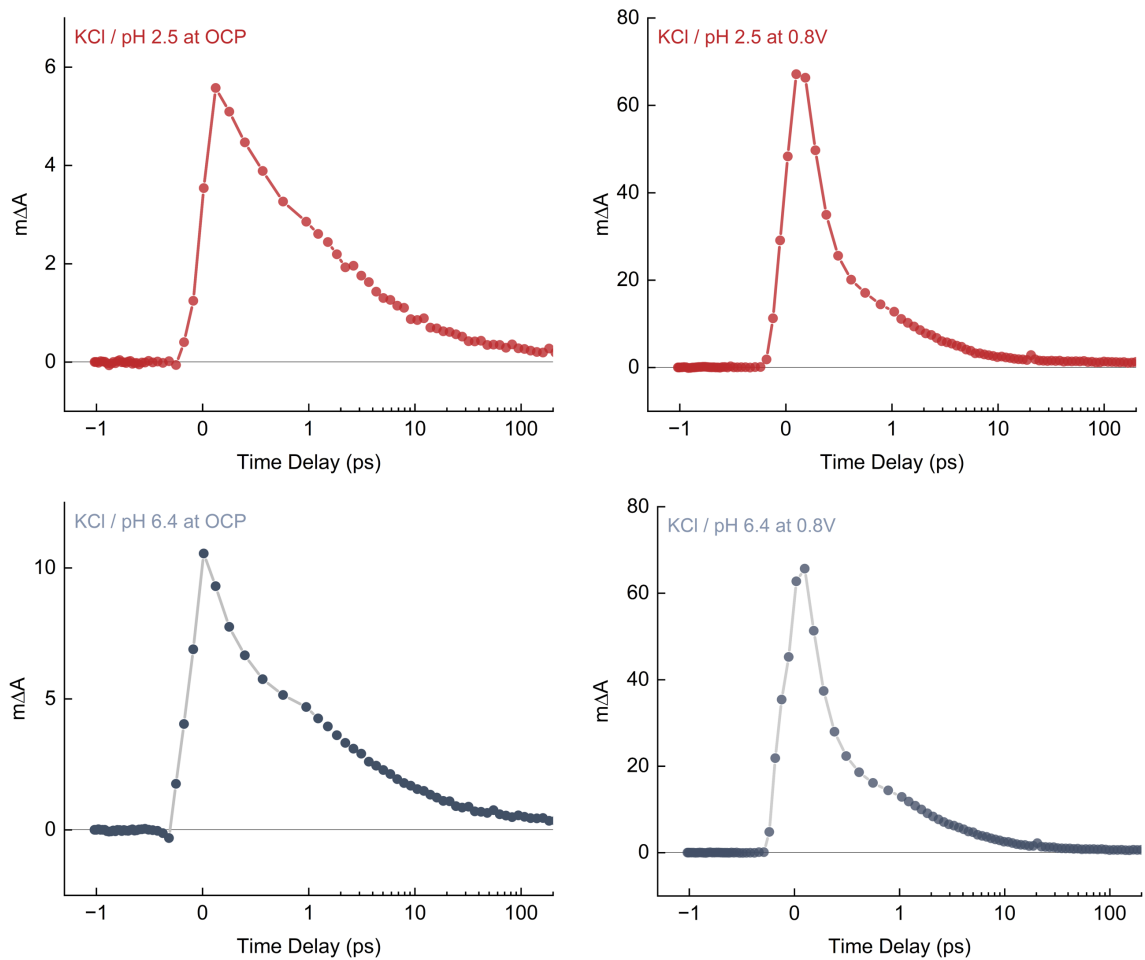

**Figure S13:** Kinetic trace at λ_obs = 1000 nm for TAS measurements on P3CBT-P under different conditions: OCP (top left), 0.8 V (top right) under pH acidic KCl conditions; and OCP (bottom left), and 0.8 V under pH neutral conditions (bottom right). Traces show the rate of decay for the TAS signal associated with the (polaron population) with much of the signal lost by ~10 ps after excitation at 850 nm.

**Discussion 2: quantify fixed-charge density vs pH and identify dominating ionic species.**

Open circuit potential (OCP) reflects the combined influence of ionic diffusion and interfacial potential, and it related to the diffusion potential by the equation: $E_{diffusion}$ = OCP − $E_{Redox}$. $E_{redox}$ term accounts for the unequal potential drop at the electrode-electrolyte interface.[13] To estimate the $E_{diffusion}$ term, we first measured the $E_{redox}$ term, which captures the unequal potential drop at the electrode-electrolyte interface, using bare ITO electrolyte across various pH conditions in 0.1 M KCl. Subsequently, the OCP values were recorded using polymer coated ITO at the start of each CV cycle for 1 minute until stabilization. The calculated values are compiled in Table S1. The polarity of $E_{diffusion}$ indicates a dominant ionic species: negative values (< 0) suggest preferential anion diffusion, while the positive values (> 0) reflect cation dominance.

Unlike previously reported systems based on charge-neutral OMIECs, carboxylic acid functionalized polymers are conjugated polyelectrolytes, exhibiting pH-dependent fixed charge due to the ionizable nature of the side chains. And our main results have shown evidence of dual ion interaction during (de)doping cycles. At low pH (2.5), the polymer is mostly in its protonated (-COOH) from, resulting in minimal fixed charge and weak Donnan exclusion of anions. As pH increases, deprotonation leads to the accumulation of fixed negative charges (-COO-), progressively favoring cation uptake, and excluding anions. This transition is directly reflected in the measured diffusion potential as shown in Table, which shifts from negative values at low and neutral pH – indicative of anion-dominated ion gradient – to near-zero or even positive values at high pH, indicating a switch to cation-dominated diffusion.

*The anion contents within the film arises from both deprotonated carboxylates (COO-) and chloride ions in the electrolyte. At low pH, despite fewer COO- group due to higher protonation, stronger anion dominance is observed – likely due to increased Cl⁻ uptake resulting from weaker Donnan exclusion and reduced cation demand.

**Notably, the different behavior between P3CBT-P and P3CBT likely originate from processing factor, since P3CBT-P is derived from salt-form (COOK) polymer followed by acid treatment, while P3CBT is directly dissolved from its protonated form in DMSO. Previous work shows that P3CBT-P retains residual potassium, which can shift the Donnan potential and modify ion compensation. An additional contributing factor may be an intrinsic difference in acid content: P3CBT-P likely with a higher acid content (lower pKa), shows a full sign reversal in diffusion term by pH 9.7, while P3CBT only trend upward (higher pKa) – this behavior aligns with the expected pKa effect, as a more acidic polymer would indeed deprotonate more at a given pH, leading to a stronger Donnan potential and earlier crossover to the cation-dominance behavior. However, this hypothesis requires further experimental validation

**Table S4:** Qualitative interpretation of Open-Circuit Potentials in determining preference in OCP values for P3CBT-P and P3CBT in pH varied KCl electrolyte.

| **P3CBT-P** | *pH 2.5* | *pH 6.4* | *pH 9.7* |
|---|---|---|---|
| OCP (V) | 320 mV | 200 mV | 205 mV |
| $E_{Redox}$ (V) | 467 mV | 333 mV | 128 mV |
| $E_{Diffusion}$ (V) | **-147 mV** | **-133 mV** | **77 mV** |

| **P3CBT** | *pH 2.5* | *pH 6.4* | *pH 9.7* |
|---|---|---|---|
| OCP (V) | 272 mV | 188 mV | 67 mV |
| $E_{Redox}$ (V) | 467 mV | 333 mV | 128 mV |
| $E_{Diffusion}$ (V) | **-195 mV** | **-145 mV** | **-61 mV** |

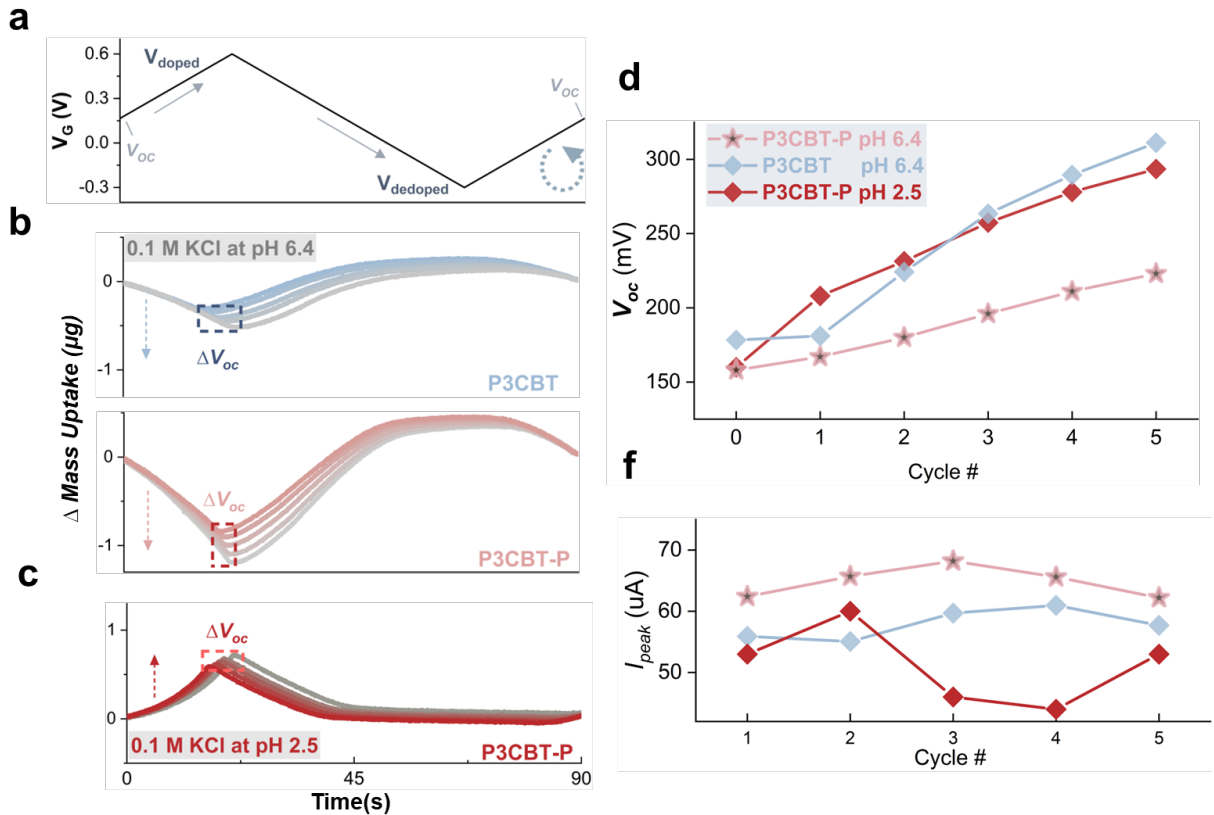

**Figure S14: Open-circuit potential (OCP or $V_{oc}$) regulates ion uptakes in carboxylated mixed conductors.** a) Experimental protocol: starting from OCP, the potential was cycled to 0.6 V, then to -0.3 V, and returned to OCP, repeated for 5 cycles with 20 second break periods between each cycle. b) Measurement for both P3CBT and P3CBT-P were conducted in 0.1 M KCl at pH neutral conditions. c) Measurement for P3CBT-P in 0.1 M KCl under acidic conditions. Since the OCP in acidic environments is typically higher than under neutral conditions, an initial preconditioning step was applied to lower the OCP to ~150 mV to enable a meaningful comparison. d) evolution of OCP, and e) peak current across cycles for all samples. The results show that both P3CBT in pH-neutral and P3CBT-P in pH-acidic conditions exhibit larger OCP drift across cycles, in contrast to the more stable behaviors of P3CBT-P at pH neutral. These findings suggest that strong cation-polymer interaction plays a functional role in stabilizing the OCP by buffering potential fluctuation during electrochemical cycling. The OCP may serve as an indicator of cation uptake and the extent of hydrophilic polymer-electrolyte interactions in carboxylated mixed conductors.

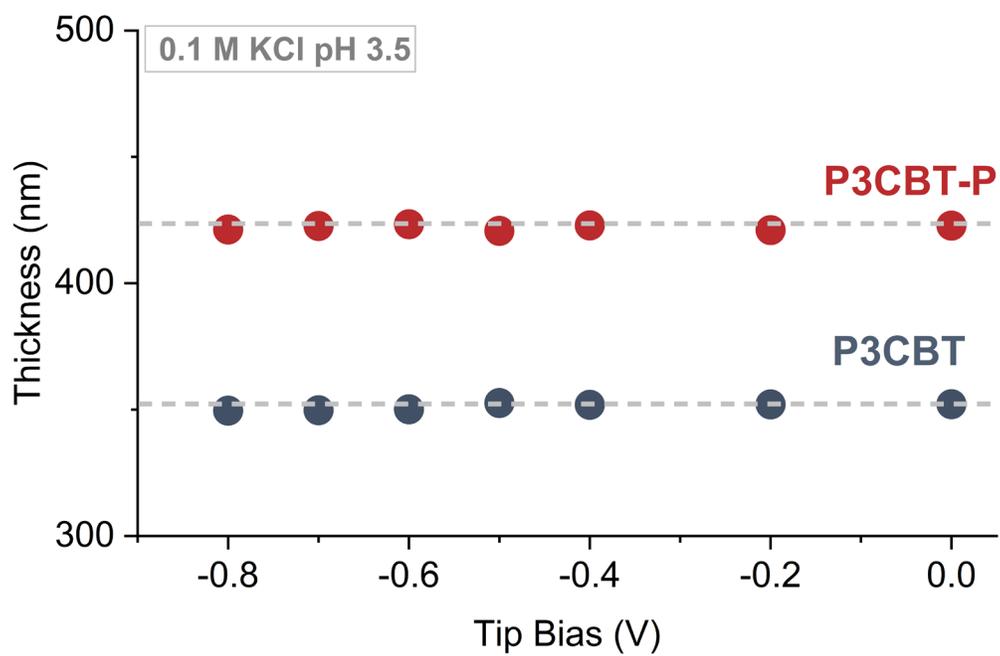

**Figure S15:** *In situ* AFM measurements of P3CBT-P and P3CBT during electrochemical doping at pH 3.5, showing negligible changes in absolute film thickness.

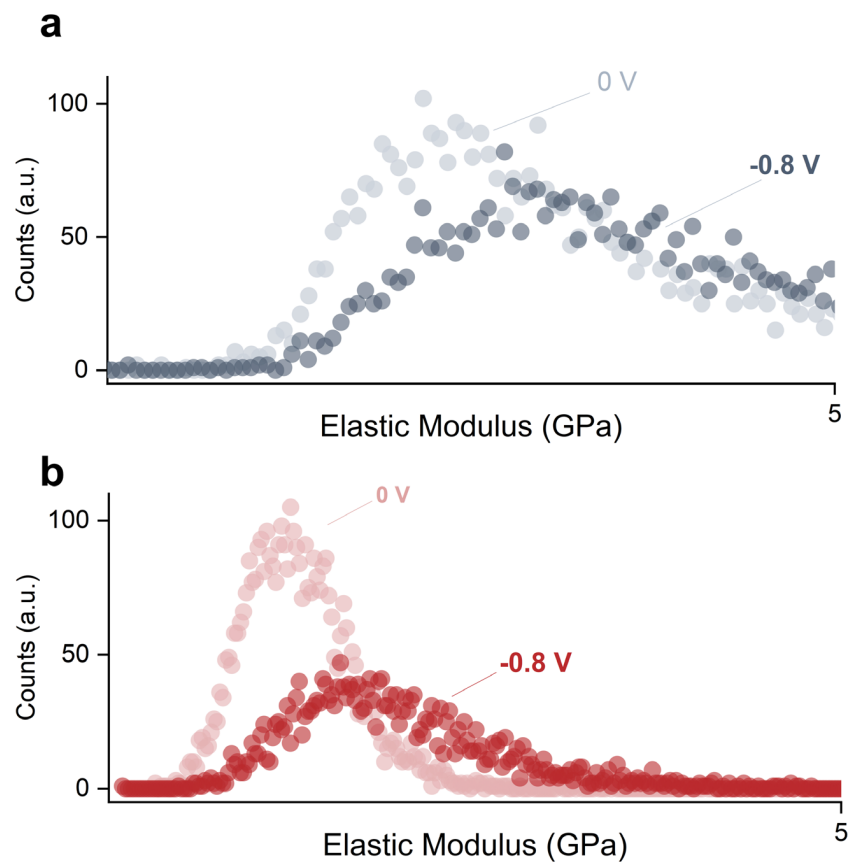

**Figure S16:** Operando force mapping revealing elastic modulus increase in a) P3CBT, and b) P3CBT-P during electrochemical doping at pH 3.5.